\documentclass[12pt]{iopart}

\usepackage{graphicx,xcolor}
\expandafter\let\csname equation*\endcsname\relax
\expandafter\let\csname endequation*\endcsname\relax
\usepackage{amsmath, amsthm, amssymb}
\UseRawInputEncoding

\usepackage{float}  %???????????
\usepackage{subfigure}  %?????????????

\begin{document}
	
	\title{Weak and reversed magnetic shear effects on
		internal kink and fishbone modes}
	
	\author{Weikang Cai}
	
	\address{
	State Key Laboratory of Advanced Electromagnetic Technology, \\International Joint Research Laboratory of Magnetic Confinement Fusion and Plasma Physics, School of Physics,
	\\	Huazhong University of Science and Technology, Wuhan, Hubei, 430074,
	China}
	
	\author{Ping Zhu*}
	
	\address{State Key Laboratory of Advanced Electromagnetic Technology, \\International Joint Research Laboratory of Magnetic Confinement Fusion and Plasma Physics, School of Electrical and Electronic Engineering,
		\\	Huazhong University of Science and Technology, Wuhan, Hubei, 430074,
		China;
		~\\
		Department of Nuclear Engineering and Engineering Physics, 
		\\University of Wisconsin-Madison, Madison,
		Wisconsin, 53706, United States of America}
	\ead{zhup@hust.edu.cn}
	\author{Zhi Zhang}
	\address{
		State Key Laboratory of Advanced Electromagnetic Technology, \\International Joint Research Laboratory of Magnetic Confinement Fusion and Plasma Physics, School of Physics,
		\\	Huazhong University of Science and Technology, Wuhan, Hubei, 430074,
		China}
		
	\author{Shiwei Xue}
	\address{State Key Laboratory of Advanced Electromagnetic Technology, \\International Joint Research Laboratory of Magnetic Confinement Fusion and Plasma Physics, School of Electrical and Electronic Engineering,
		\\	Huazhong University of Science and Technology, Wuhan, Hubei, 430074,
		China}
	
	\author{Sui Wan}
	\address{State Key Laboratory of Advanced Electromagnetic Technology, \\International Joint Research Laboratory of Magnetic Confinement Fusion and Plasma Physics, School of Electrical and Electronic Engineering,
		\\	Huazhong University of Science and Technology, Wuhan, Hubei, 430074,
		China}
		
	\vspace{10pt}
	\begin{indented}
		\item[]February 7th, 2026
	\end{indented}
	
\newpage

	\begin{abstract}
		
		Advanced tokamak scenarios often feature weak or reversed magnetic shear
		configurations. In this study, the hybrid kinetic-MHD model implemented in the
		NIMROD code is used to investigate the effects of reversed magnetic shear on internal
		kink and fishbone mode in a circular shaped limiter tokamak. In absence of energetic
		particles (EPs), the mode growth rate initially increases and then decreases as the
		magnetic shear changes from positive to negative, indicating stabilizing effects of
		the reversed magnetic shear on the internal kink mode. In presence of EPs,  when the
		reversed magnetic shear region is sufficiently narrow, the transition from internal kink/fishbone
		modes to double kink/fishbone modes takes place, and the stabilizing effects of the
		reversed magnetic shear can significantly dominate the destabilization of EPs. For non-resonant modes, the EP beta fraction $\beta\rm_f$ for excitation increases with $q_{\rm min}$, concurrent with progressively lower growth rates in non-resonant fishbone modes. When the equilibrium profile has an internal transport barrier (ITB), broader ITB widths suppress internal kink modes more effectively, whereas steeper temperature gradients strengthen EP stabilization. 
		
	\end{abstract}
	
	\vspace{2pc}
	
	%\submitto{\PST}
	%\maketitle
	\section{Introduction}
	
	The internal kink instability imposes constraints on the maximum plasma current density in the core region of a tokamak. The energetic particles (EPs) from auxiliary
	heating and fusion reactions can strongly interact with the internal kink mode \cite{shafranov70a,rosenbluth73a,bussac75a} and excite the fishbone mode. Since the initial observation of fishbone modes on
	PDX \cite{mcguire83a} during high-power Neutral Beam Injection (NBI) in 1980s, these modes have been observed on other tokamaks, including DIII-D \cite{wong00a}, JET \cite{nave91a}, HL-2A \cite{chen10a}, and EAST \cite{xu15a}, among others, in presence of various forms of auxiliary heating, such
	as perpendicular/parallel NBI, Electron Cyclotron Resonant Heating (ECRH), or Ion Cyclotron Resonant Heating (ICRH). Theoretical studies indicate that fishbone modes
	are triggered by the resonant interaction between the trapped EPs and the internal kink modes, as well as the diamagnetic dissipation from the background plasmas \cite{chen84a, coppi86a} or
	the resonance between the circulating EPs and the internal kink modes, along with the effects of finite radial particle-orbit excursion \cite{betti93a,wang01a}. Based on EAST-like parameters,
	using the kinetic-MHD hybrid code M3D-K for simulations of the fishbone instabilities, Shen et al show that the dual resonant fishbone (DRF) is excited by the trapped beam
	ions when the fast ion pressure exceeds a critical value, and the mode structure of DRF exhibits splitting radial structure due to double $q = 1$ surfaces \cite{shen20a}.
	
	During the NBI heating of plasmas with flat central safety factor profile, long-lived modes (LLMs) are observed on HL-2A that can last for several hundred milliseconds, and the mode frequency is approximately proportional to the toroidal mode number $n$ \cite{wei13a}. The LLMs can be suppressed by ECRH or SMBI (supersonic molecular beam injection) which is related to changing of $q$ profile and pressure gradient\cite{wei13a}. The excitation of fishbone modes in tokamak plasmas with positive magnetic shear depends on $q^{\prime}=dq/dr$ at $q = 1$ surface\cite{bussac75a,hu06a}. However, in the case of reversed magnetic shear regimes, conventional fishbone analysis may not be sufficient to explain the fast ion driven mode \cite{zonca07a,zonca14a}. When the safety factor profile $q_0$ is close to unity, the fast ion-driven mode can be observed even with a non-monotonic q-profile, as reported in experiments conducted on DIII-D \cite{wong00a}, NSTX \cite{wang13a}, and MAST \cite{chapman10a}. The growth rate of the kink mode with the reversed magnetic shear given is also valid in $q_{\rm min}>1$ and $q_{s}^{\prime}=0$ cases\cite{hastie87a}.
	
	The impact of safety factor profile, particularly in presence of very weak magnetic shear, on the $m/n = 1$ mode triggered by energetic ions in tokamak plasmas has been investigated through numerical simulations and theoretical analysis \cite{li15a, meng15a, wang15a, zou22a}. In presence of reversed magnetic shear, a double fishbone mode (DFM) can arise when $q_{\rm min} < 1$, whose mode structure is distinct from that of the conventional double kink mode. The frequency of DFM is found to be in close proximity to half of the average EP precession frequency \cite{meng15a}. Solving an analytical dispersion relation developed for non-resonant fishbone (NRF) with reversed magnetic shear configuration, Wang et al find that the growth rate with $q_{\rm min} > 1$ depends on the fast ion beta $\beta_h$ in a power law of $\sim \beta_h^{2/3}$, different from that of $\sim \beta_h$ in a conventional positive magnetic shear configuration \cite{wang15a}. Recent 3D resistive MHD simulations using the M3D-C1 code have also identified a dominant non-resonant $(1,1)$ infernal mode in spherical tokamak plasmas when $q_{\rm min}$ approaches unity from above ($q_{\rm min} \sim 1.1-1.2$), which can lead to a nonlinear flattening of the core current and pressure profiles \cite{jardin2025resistive}. Based on the equilibrium with flat q-profile in core plasma region generated from the LLM experiments on HL-2A, Zou et al show that the EP-driven fishbone mode frequency is almost proportional to the toroidal mode number $n$ even in absence of any equilibrium toroidal flow, using the hybrid kinetic-MHD model implemented in the NIMROD code \cite{zou22a}.
	
	Despite this progress, a consistent understanding on the effects of reversed magnetic shear on internal kink and fishbone modes based on both theory and simulation
	remains desirable. In this study, we investigate the influence of continuously various $q$-profiles on the instability induced by energetic particles (EPs), using simulation based on the hybrid kinetic-MHD model implemented in the NIMROD code \cite{kim08a, sovinec04a, brennan12a}, and in the context of previous theoretical formulation. Both simulation and theory have been able to demonstrate the dominant stabilizing effects of the reversed magnetic shear over the destabilizing effects of EPs on internal kink mode.
	
	The remainder of this paper is structured as follows. Section 2 provides a brief review of the simulation model. In section 3 we specify the simulation setup. In
	section 4, we report the main results about the effects of magnetic shear and EP $\beta$ fraction on the kink/fishbone mode. Finally, in section 5, we summarize and discuss the findings.
	\section{Simulation model}
	
	This study is based on a set of hybrid kinetic-MHD equations implemented in the NIMROD code \cite{kim08a, sovinec04a}
	\begin{gather}
			\frac{\partial \rho}{\partial t}+\nabla \cdot(\rho \mathbf{V})=0, \\
			\rho\left(\frac{\partial \mathbf{V}}{\partial t}+\mathbf{V} \cdot \nabla \mathbf{V}\right)=\mathbf{J} \times \mathbf{B}-\nabla p_b-\nabla \cdot P_h, \\
			\frac{n}{\Gamma-1}\left(\frac{\partial T}{\partial t}+\mathbf{V} \cdot \nabla T\right)=-p\nabla \cdot V-\nabla \cdot \mathbf{q}+Q, \\
			\frac{\partial \mathbf{B}}{\partial t}=-\nabla \times \mathbf{E}, \\
			\nabla \times \mathbf{B}=\mu_0 \mathbf{J} ,\\
			\mathbf{E}=-\mathbf{V} \times \mathbf{B},
	\end{gather}
	where $\rho, \mathbf{V}, \mathbf{J}, p_b, n$ and $T$ are the mass density, center of mass velocity, current density, plasma pressure, number density, and temperature of the main species plasma, $\mathbf{E}$ and $\mathbf{B}$ are the electric and magnetic fields, $\Gamma$ and $\mu_0$ are the specific heat ratio and the vacuum permeability, respectively. $\mathbf{q}$ and $Q$ are the heat flux and the heating source terms, which are not considered for the modes we study here. By adding the pressure tensor of energetic particles to the momentum equation, which is defined as
	\begin{equation}
		P_h=m_h \int \Delta \mathbf{v}_{\mathbf{h}} \Delta \mathbf{v}_{\mathbf{h}} f_h d^3 \mathbf{v}_{\mathbf{h}},
	\end{equation}
	the effects of energetic particles can be included, where $\Delta \mathbf{v}_{\mathbf{h}}=\mathbf{v}_{\mathbf{h}}-\mathbf{V}_{\mathbf{h}}$, and $m_h, \mathbf{v}_{\mathbf{h}}, f_h, \mathbf{V}_{\mathbf{h}}$ are the mass, velocity, distribution function, and the center of mass velocity of EPs, respectively.
	\section{Simulation setup}
	
\begin{figure}[h]
	\centering
	\begin{minipage}{0.55\linewidth} 
		\raggedleft 
		
		\subfigure{
			\includegraphics[width=0.97\linewidth]{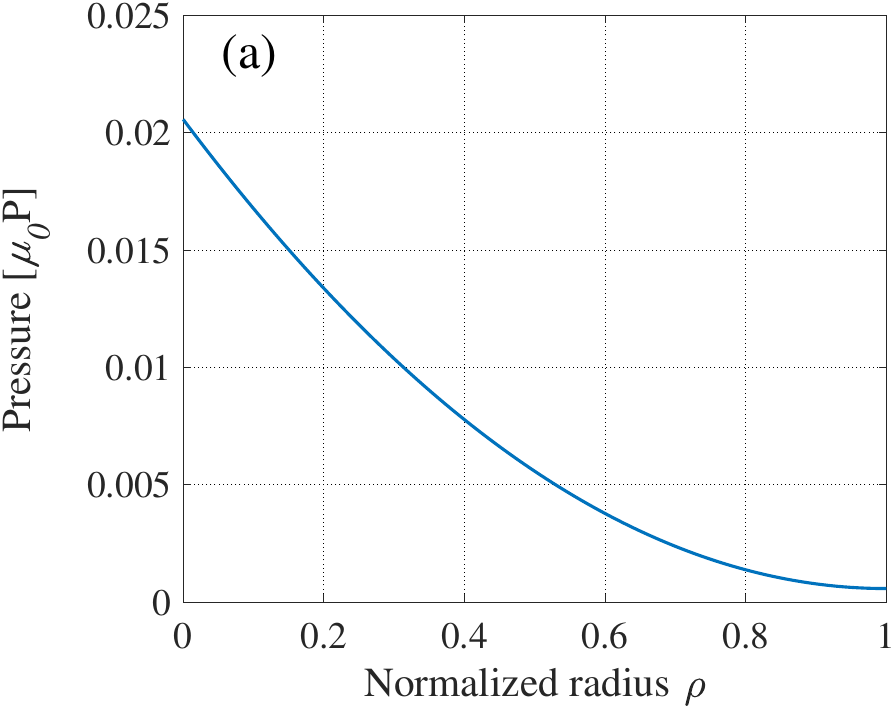} 
		
			\label{fig:fig1.1}
		}
		\\
		\subfigure{
			\includegraphics[width=0.97\linewidth]{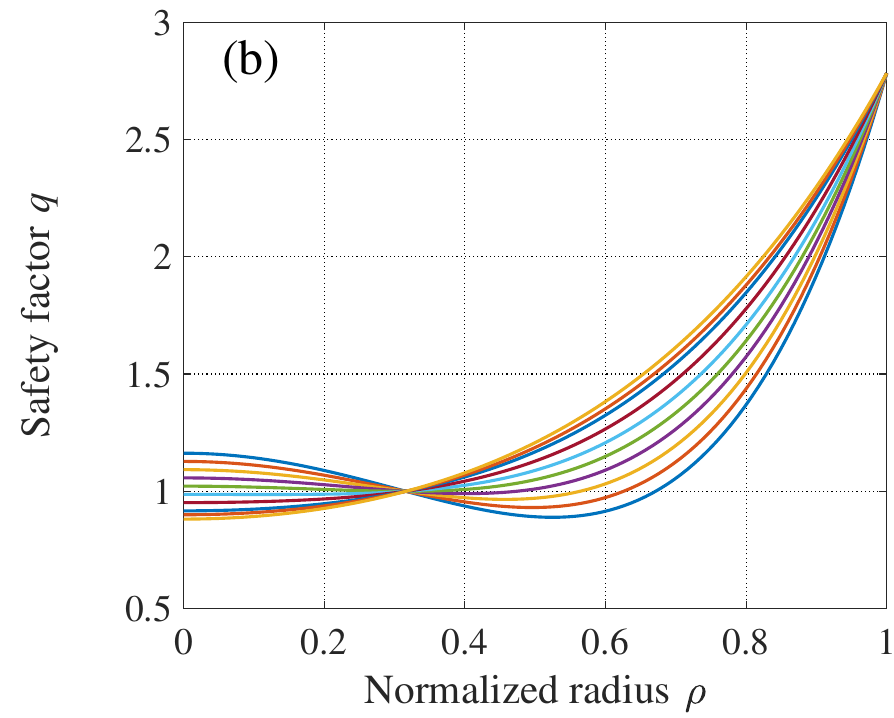}
			\label{fig:fig1.2}
		}
		\\
		\subfigure{
			\includegraphics[width=0.9\linewidth]{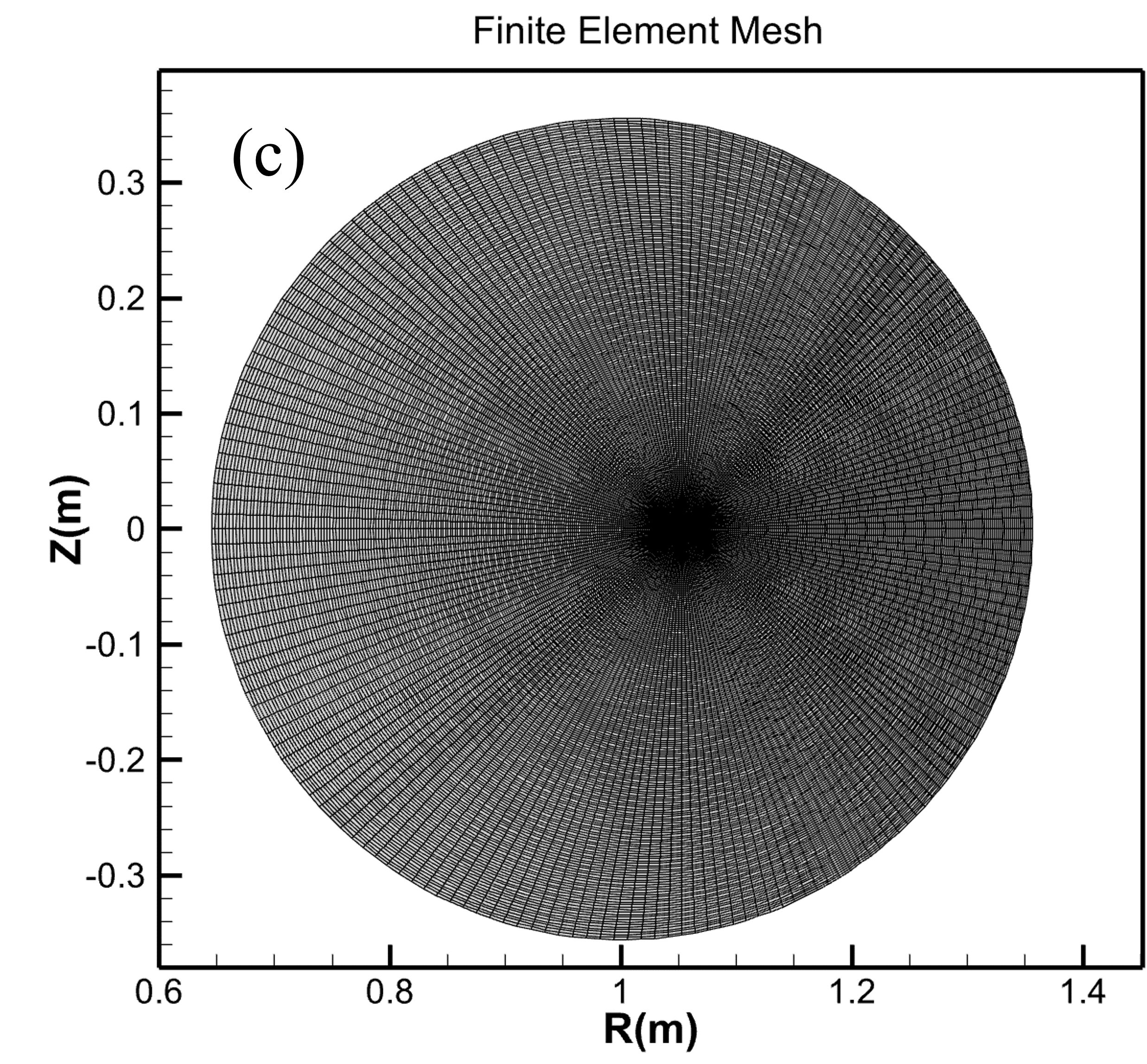}
			\label{fig:fig1.3}
		}
	\end{minipage}
	\caption{(a) Equilibrium pressure profile and (b) $q$-profiles as functions of normalized minor radius $\rho$ with various magnetic shear $\hat{s}$ at $q=1$ surface. (c) Equilibrium magnetic flux-aligned mesh for poloidal plane used in NIMROD simulations.}
	\label{fig:fig1}
\end{figure}

	 A circular-shaped tokamak equilibrium with the pressure profile $q(\psi)=a(\psi-c)^2+b(\psi-c)(\psi-1)+1$ and the safety factor profile $p(\psi)=p_0\left(\psi^2-2 \psi+1\right)$ is generated using the CHEASE code, where $\psi$ represents the normalized poloidal magnetic flux and a magnetic flux-aligned mesh is adopted (figure \ref{fig:fig1}). The internal kink mode is located within a region $0<\sqrt{\psi / \psi_0}<0.6$ where the safety factor $q$-profile is below unity and the magnetic shear is weak [figure \ref{fig:fig1.2}]. Here, $\psi$ denotes the poloidal magnetic flux and $\psi_0$ represents the total poloidal magnetic flux inside the last closed flux surface.
	
	Initially the energetic particles are assumed to follow the slowing-down distribution \cite{kim08a},
	
	\begin{equation}
		f_0=\frac{P_0 \exp \left(\frac{P_c}{\psi_n}\right)}{\varepsilon^{3 / 2}+\varepsilon_c^{3 / 2}},
	\end{equation}
	where $P_0$ is the normalization constant, $P_\zeta=g \rho_{\|}-\psi_p$ is the canonical toroidal momentum, $g=R B_\phi, \rho_{\|}=m v_{\|} / q B, \psi_p$ is the poloidal flux, $\psi_n=c \psi_0, \psi_0$ is the total poloidal flux and the parameter $c$ is used to match the spatial profile of the equilibrium, $\varepsilon$ is the EP energy, and $\varepsilon_c$ is the critical slowing-down energy \cite{goldston20book}
	
	\begin{equation}
		\varepsilon_c=\left(\frac{3}{4}\right)^{2 / 3}\left(\frac{\pi m_i}{m_e}\right)^{1 / 3} T_e,
	\end{equation}
	with $m_i$ being the ion mass, $m_e$ the electron mass, and $T_e$ the electron temperature. In our linear simulation, we set resistivity $\eta=0$, and a total $2 \times 10^6$ simulation particles are prescribed in the poloidal plane with $64 \times 64$ bicubic finite elements. The remaining parameters are set up as follows: major radius $R_0=1.0 \mathrm{~m}$, aspect ratio $R_0/a=2.8$, magnetic axis $R_m=1.06 \mathrm{~m}$,  maximum toroidal magnetic $B_\mathrm{max}=1.0 \mathrm{~T}$, and the number density is uniform in the radial direction with $n_e=2.489 \times 10^{19} \mathrm{~m}^{-3}$ \cite{kim08a}.

	\section{Calculation results}
	\subsection{Magnetic shear effects in absence of EPs}
	
	In figure 1(b), we vary the value of $b$ for the same values of $a$ and $c$, thereby altering the magnetic shear $\hat{s}$ from -0.8 to 0.8 , with the location of the $q=1$ surface fixed at $\psi_s=0.1$.
	
	In absence of EPs, the $(1,1)$ internal mode growth rate first increases as the positive magnetic shear decreases to zero and becomes reversed, reaching a maximum value near $\hat{s}=-0.2$, and then starts to decrease as the magnetic shear reversal continues. Such shear effects may be understood from the previous analytical theory, where the growth rate of the mode in positive magnetic shear region can be expressed as  \cite{li15a}
	\begin{equation}\label{eq:10}
		\frac{\gamma}{\omega_A}=-\frac{1}{\sqrt{3} \hat{s}} \delta \widehat{W}_{\rm MHD},
	\end{equation}
	where $\omega_A=V_A / q R_0=V_A / R_0$ is the Alfvén frequency on the $q=1$ surface, $\hat{s}=r_s q^{\prime} / q$ is the magnetic shear at $q=1$ surface, $r_s$ is the radius of the $q=1$ surface, and $\delta \widehat{W}_{\rm MHD}$ can be written in the form of
	
	\begin{equation}\label{eq:11}
		\delta \widehat{W}_{\rm MHD} \approx 3 \pi \delta q \varepsilon_1^2\left(\beta_{p c}^2-\beta_p^2\right),
	\end{equation}
	where the poloidal plasma beta
	
	\begin{equation}\label{eq:12}
		\beta_p=-\left(8 \pi / B_0^2\right) \int_0^{r_{\star}}\left(r / r_s\right)^2(d p / d r) d r,
	\end{equation}
	$\varepsilon_1=r_s / R_0$, and $\delta q=1-q(r=0)=1-q_0$. For flattened $q$-profiles with $q_0<1$ and a finite $\delta q$, one finds a threshold of plasma beta, i.e., $\beta_{p c}=\left[\frac{13(4-v)}{48(4+v)}\right]^{1 / 2}$, for a parabolic $q$-profile, $v=2$. From Eq.~\eqref{eq:10}, it can be seen that when the location of the resonant surface and the $p$-profile are fixed, as shown in figure $1(\mathrm{a})$, $\delta \widehat{W}_{\rm MHD}$ is mainly affected by $\delta q$,  which indicates that the magnetic shear $\hat{s}$ is stabilizing whereas $\delta q$ is destabilizing on the mode. In positive magnetic shear region, it can be concluded that when $\hat{s}<0.72$, the increase of magnetic shear plays a dominant role in enhancing the stabilizing effects on the mode. When $\hat{s}>0.72$, the increase of $\delta q$ dominates the enhanced destabilizing effects on the mode and increasing the growth rate, as shown in figure 2(a). Previously, only the case where the growth rate decreases with the increase of magnetic shear is reported \cite{li15a}. This may be because the magnetic shear at the resonant surface is not large enough to reveal the destabilizing effects of $\delta q$ on the mode.
	
	In the weak reversed magnetic shear region, the mode growth rate first increases and then decreases with decreasing magnetic shear, reaching a maximum value near $\hat{s}=-0.2$. As shown in figure 2(b), the growth rate increases first and then decreases as $q_{\text {min }}$ increases, which is consistent with the analytical result in \cite{li15a}
	\begin{equation}\label{eq:13}
		\bar{\gamma}\left[(\bar{\gamma}+\Delta q) r_{\min }^2 q^{\prime \prime}\right]^{1 / 2}=-\delta \widehat{W}_{\rm MHD},
	\end{equation}
	where $\bar{\gamma}=\left[3 \gamma^2 / \omega_A^2+(\Delta q)^2\right]^{1 / 2}, q_{\min }=q\left(r_{\min }\right), \Delta q=q_{\min }-1, r_{\min }^2 q^{\prime \prime}=r_{\min }^2 d^2 q / d r^2$ at $q=q\left(r_{\min }\right)$ surface, and $\delta \widehat{W}_{\rm MHD}$ is given in Eq. ~\eqref{eq:11}. In comparison with the $q$-profiles in \cite{li15a}, for a fixed resonance surface, in this study the change in the magnitude of $r_{\min }^2 q^{\prime \prime}$ near the $q=q_{\min }$ surface is not significant. Therefore, the reason for the growth rate variation with $q_{\min }$ is mainly due to the variation in $\Delta q$. The growth rate of the $q_{\min }<1$ profile rises with $q_{\min }$, as a result of the enhanced coupling between the two resonant surfaces as $q_{\min }$ increases. When the $q_{\min }$ surface approaches the $q=1$ surface, the mode growth rate decreases, likely because the mode becomes less resonant.
	
	The plasma pressure perturbation contour indicates that the mode structure contracts as $\hat{s}$ increase (figure 3). For the positive magnetic shear, the mode structure is a single kink mode within the $q<1$ region; for the reversed magnetic shear, the mode structure is that of a double kink mode between two $q=1$ surfaces. The mode structures in figure 3(a) and (b) are highly consistent with those obtained using analytical methods in \cite{meng15a}. When the two resonant surfaces are relatively far apart, the double kink modes structure appears more prominent, with the first layer of modes located within the first resonant surface and the second layer of modes between the two resonant surfaces. As the two resonant surfaces draw closer together, the layered mode structure becomes less obvious.
	
	\subsection{Magnetic shear effects in presence of EPs}
	
	In presence of EPs, when $\beta\rm_f$ is low, e.g. $\beta\rm_f=0.1$, the growth rates are reduced, but the overall trend is similar to the case without EPs [figure 4(a)]. When $\beta\rm_f$ increases, e.g. $\beta\rm_f=0.5$, the growth rate increases relative to the case without EPs when $\hat{s} > -0.25$, which is attributed to the excitation of fishbone modes. In contrast, for $\hat{s} < -0.25$, the growth rate decreases, and increasing magnetic shear further reduces the growth rate, demonstrating its stabilizing effect \cite{li15a}.
	
	In the case where the magnetic shear is between -0.2 and 0.2 (figure 5), we observe that the mode growth rate and real frequency exhibit a similar trend. Specifically, the growth rate decreases initially, and then increases with an increase in $\beta\rm_f$, whereas the real frequency increases with $\beta\rm_f$. For the case with $\hat{s}=-0.6,-0.8$ (figure 6), it is observed that the presence of energetic particles can lead to a reduction in the growth rate of the mode, indicating stabilizing effects. The growth rate of the mode can be expressed as \cite{meng15a}
	\begin{equation}\label{eq:14}
		\frac{\gamma}{\omega_A}=\frac{\left|\operatorname{Re}\left(\delta \widehat{W}_K\right)\right|}{\sqrt{3}\left[(1-M)^2 \hat{s}_1+M^2\left(r_2 / r_1\right)^2 \hat{s}_2\right]},
	\end{equation}
	where $\delta \widehat{W}_K$ is the kinetic energy of trapped fast particles, $r_1$ and $r_2$ are the positions of the two resonant surfaces, $M$ denotes the ratio of displacement amplitudes at the inner and outer $q=1$ resonant surfaces, and $\hat{s}_1$ and $\hat{s}_2$ are the magnetic shears of the two resonant surfaces. When $\hat{s}_1$ is sufficiently negative, $(1-M)^2 \hat{s}_1+M^2\left(r_2 / r_1\right)^2 \hat{s}_2<0$, which leads to a stronger stabilizing effects on the mode when the energetic particles fraction increases. 
	
	Additionally, the real frequency of the mode still shows a monotonic increase with increasing $\beta\rm_f$, even for $\hat{s}=-0.6$. However, for the case with $\hat{s}=-0.8$, the real frequency no longer shows a monotonic increase with increasing $\beta\rm_f$. The plasma pressure perturbation contours (figure 7) indicate that the mode structure transitions from a double kink mode to a double fishbone mode as $\beta\rm_f$ increases.

	\subsection{Non-resonant modes in presence of EPs}

	The growth rate of the non-resonant mode when $q_{\rm min}>1$ can be expressed as \cite{Wang_2014}

	\begin{equation}\label{eq:15}
	\gamma / \omega_A=\frac{1}{\sqrt{3} S^{2 / 3}} \operatorname{Re}\left[\left(-\delta \widehat{W}_{\rm MHD}-\delta \widehat{W}_K\right)^{2 / 3}\right]-\Delta q /3 \sqrt{3},
	\end{equation}
	where $S=\sqrt{r_s^2 q^{''}_s},\Delta q=q_{\rm min}-1$. Equation (\ref{eq:15}) demonstrates that $\Delta q$ stabilizes the mode, whereas $\beta_{\rm f}$ destabilizes it. When $\Delta q$ decreases below 0.02, the term $\Delta q/3\sqrt{3}$ becomes negligible. The mode growth rate consequently reduces to that of the resonant fishbone mode derived in \cite{li15a}. Increasing $\Delta q$ raises the critical $\beta_{\rm f}$ required for non-resonant mode excitation and simultaneously lowers the growth rate.  
	
	The above theory prediction on the non-resonant mode is verified in our simulations. Given that the typical experimental uncertainty of the safety factor profile is approximately $5\% \sim 10\%$ \cite{Soltwisch_1986, Levinton_1993}, simulations are performed with $q_{\rm min}$ varying from $1.01$ to $1.20$. For $q_{\rm min} \geq 1$, the growth rate of non-resonant internal kink exhibits similarities to the resonant modes (figure 8(b)). Increasing $\beta\rm_f$ drives a transition from the internal kink to the fishbone mode. Resonant internal kink modes are absent for $q_{\rm min}>1.02$. Above critical $\beta\rm_f$ thresholds, only non-resonant fishbone modes emerge, with growth rates decreasing as $q_{\rm min}$ rises, in agreement with the $\Delta q$ stabilization indicated in Eq.(\ref{eq:15}). A similar dependency on the safety factor profile is also found in resistive MHD simulations, where a dominant non-resonant $(1,1)$ infernal mode is excited as $q_{\rm min}$ decreases towards unity, further confirming the destabilizing effect of reducing $\Delta q$ \cite{jardin2025resistive}.

	\subsection{Effects due to internal transport barrier profile}
	To examine the effects from an internal transport barrier (ITB), the following equilibrium temperature profile is adopted
	
	\begin{equation}\label{eq:hyperbolic}
		T = A \frac{(1 + \alpha z) e^{z} - (1 - \beta z) e^{-z}}{e^{z} + e^{-z}} + T_0,
	\end{equation}
	where  $z = (\rho_\mathrm{ITB} - \rho)/w_{\mathrm{ITB}}$, $\rho$ is the normalized radial coordinate, $\rho_\mathrm{ITB}$ is the ITB location, and $w_{\mathrm{ITB}}$ is the ITB width. \(A\), \(\alpha\), \(\beta\), and \(T_0\) are additional profile shaping coefficients. In the simulations, we selected $\rho_\mathrm{ITB} = 0.35$, \(\text{$w_{\mathrm{ITB}}$} = 0.14\), \(A = 0.45\), \(\alpha = 0.06\), \(\beta = 0.3\), and \(T_0 = 1.45\) (figure 9).
	
	Varying the magnetic shear $\hat{s}$ from -0.8 to 0.8, the variation of mode growth rate with magnetic shear remains similar to the case in absence of an ITB (Figure 10). For the $q$ profile with reversed magnetic shear ($\hat{s} = -0.6$), the growth rate of the internal kink mode decreases as the ITB becomes wider. As the ITB becomes steeper (figure 11), the stabilizing effect of energetic particles become more pronounced (figure 12). This finding agrees with the numerical simulation results reported by Ge et al \cite{Ge2023}. 
	
	\section{Conclusions and discussions}
	
	Kinetic-MHD simulations using the NIMROD code are performed to investigate the impact of weak and reversed magnetic shear on
	the internal kink mode in tokamaks. For the resonant (1,1) mode, its growth rate increases as the magnetic shear at $q=1$ surface reduces from positive to zero and negative, until the growth rate reaches its maximum value at a finite negative magnetic shear. Further increasing the absolute value of the negative magnetic shear at $q=1$ surface starts to stabilize the (1,1) mode with reduced growth rate. Such effects of weak and reversed magnetic shear on the (1,1) kink mode persist in presence of EPs. For non-resonant modes with $q_{\rm min}$ slightly above 1, i.e. $\Delta q=q_{\rm min}-1\lesssim 0.02$, the dynamics closely resemble those of resonant modes. When $q_{\rm min}$ rises sufficiently above 1, i.e. $\Delta q\gtrsim 0.02$, the critical $\beta_{\rm f}$ required to trigger non-resonant modes grows proportionally to $\Delta q$, and the growth rate decreases with $\Delta q$. These results agree quantitatively with previous theoretical calculations \cite{li15a}. 
		
	Future work plans on exploring the effects of the weak and reversed magnetic shear on the nonlinear internal kink and fishbone mode during sawtooth and ITB formation processes in the advanced tokamak configurations.
	
	\section*{Acknowledgements}
	
	We are grateful to Prof. Xianqu Wang and Dr. Haolong Li, as well as the support of the NIMROD team. This work is supported by the National MCF Energy R\&D Program of China under Grant No.~2019YFE03050004, the Hubei International Science and Technology Cooperation Project under Grant No. 2022EHB003, and the U.S. Department of Energy Grant No.~DE-FG02-86ER53218. The computing work in this paper is supported by the Public Service Platform of High Performance Computing by Network and Computing Center of HUST.
	\clearpage
	\section*{References}
	\bibliography{sample-1}

@article{shafranov70a,
  title={{Hydromagnetic stability of a current-carrying pinch in a strong longitudinal magnetic field}},
  author={Shafranov, Vitaly D},
  journal={Soviet Physics Technical Physics},
  volume={15},
  pages={175},
  year={1970}
}

@article{rosenbluth73a,
  title={Nonlinear properties of the internal $m= 1$ kink instability in the cylindrical tokamak},
  author={Rosenbluth, Marshall N and Dagazian, R. Y. and Rutherford, P H},
  journal={The Physics of Fluids},
  volume={16},
  number={11},
  pages={1894--1902},
  year={1973},
}

@article{bussac75a,
  title={Internal kink modes in toroidal plasmas with circular cross sections},
  author={Bussac, M N and Pellat, R and Edery, D and Soule, J L},
  journal={Physical Review Letters},
  volume={35},
  number={24},
  pages={1638},
  year={1975},
  publisher={APS}
}

@article{mcguire83a,
  title={Study of high-beta magnetohydrodynamic modes and fast-ion losses in PDX},
  author={McGuire, K and Goldston, R and Bell, M and others},
  journal={Physical Review Letters},
  volume={50},
  number={12},
  pages={891},
  year={1983},
  publisher={APS}
}

@article{wong00a,
  title={Internal kink instability during off-axis electron cyclotron current drive in the DIII-D tokamak},
  author={Wong, K L and Chu, M S and Luce, T C and others},
  journal={Physical Review Letters},
  volume={85},
  number={5},
  pages={996},
  year={2000},
  publisher={APS}
}

@article{nave91a,
  title={Fishbone activity in JET},
  author={Nave, M F F and Campbell, D J and Joffrin, E and others},
  journal={Nuclear Fusion},
  volume={31},
  number={4},
  pages={697},
  year={1991},
  publisher={IOP Publishing}
}

@article{chen10a,
  title={Features of ion and electron fishbone instabilities on HL-2A},
  author={Chen, W and Ding, X T and Liu, Yi and others},
  journal={Nuclear Fusion},
  volume={50},
  number={8},
  pages={084008},
  year={2010},
  publisher={IOP Publishing}
}

@article{xu15a,
  title={Fishbone activity in experimental advanced superconducting tokamak neutral beam injection plasma},
  author={Xu, Li Qing and Zhang, Ji Zong and Chen, Kai Yun and others},
  journal={Physics of Plasmas},
  volume={22},
  number={12},
  year={2015},
  publisher={AIP Publishing}
}

@article{chen84a,
  title={Excitation of internal kink modes by trapped energetic beam ions},
  author={Chen, Liu and White, R B and Rosenbluth, M N},
  journal={Physical Review Letters},
  volume={52},
  number={13},
  pages={1122},
  year={1984},
  publisher={APS}
}

@article{coppi86a,
  title={Theoretical model of fishbone oscillations in magnetically confined plasmas},
  author={Coppi, B and Porcelli, Francesco},
  journal={Physical Review Letters},
  volume={57},
  number={18},
  pages={2272},
  year={1986},
  publisher={APS}
}

@article{betti93a,
  title={Destabilization of the internal kink by energetic-circulating ions},
  author={Betti, R and Freidberg, J P},
  journal={Physical Review Letters},
  volume={70},
  number={22},
  pages={3428},
  year={1993},
  publisher={APS}
}

@article{wang01a,
  title={Destabilization of internal kink modes at high frequency by energetic circulating ions},
  author={Wang, Shao Jie},
  journal={Physical Review Letters},
  volume={86},
  number={23},
  pages={5286},
  year={2001},
  publisher={APS}
}

@article{shen20a,
  title={Hybrid simulation of fishbone instabilities with reversed safety factor profile},
  author={Shen, Wei and Wang, Feng and Fu, G Y and Xu, Li Qing and Ren, Zhen Zhen},
  journal={Nuclear Fusion},
  volume={60},
  number={10},
  pages={106016},
  year={2020},
  publisher={IOP Publishing}
}

@article{wei13a,
  title={Investigation of the long-lived saturated internal mode and its control on the HL-2A tokamak},
  author={Wei, Deng and Yi, Liu and Wang, X Q and others},
  journal={Nuclear Fusion},
  volume={54},
  number={1},
  pages={013010},
  year={2013},
  publisher={IOP Publishing}
}

@article{hu06a,
  title={Kinetic stability of the internal kink mode in ITER},
  author={Hu, Bo and Betti, R and Manickam, J},
  journal={Physics of Plasmas},
  volume={13},
  number={11},
  year={2006},
  publisher={AIP Publishing}
}

@article{zonca07a,
  title={Electron fishbones: theory and experimental evidence},
  author={Zonca, F and Buratti, P and Cardinali, A and others},
  journal={Nuclear Fusion},
  volume={47},
  number={11},
  pages={1588},
  year={2007},
  publisher={IOP Publishing}
}

@article{zonca14a,
  title={Theory on excitations of drift Alfv{\'e}n waves by energetic particles. II. The general fishbone-like dispersion relation},
  author={Zonca, Fulvio and Chen, Liu},
  journal={Physics of Plasmas},
  volume={21},
  number={7},
  year={2014},
  publisher={AIP Publishing}
}

@article{wang13a,
  title={Linear stability and nonlinear dynamics of the fishbone mode in spherical tokamaks},
  author={Wang, Feng and Fu, G Y and Breslau, J A and others},
  journal={Physics of Plasmas},
  volume={20},
  number={10},
  year={2013},
  publisher={AIP Publishing}
}

@article{chapman10a,
  title={Saturated ideal modes in advanced tokamak regimes in MAST},
  author={Chapman, I T and Hua, M D and Pinches, S D and others},
  journal={Nuclear Fusion},
  volume={50},
  number={4},
  pages={045007},
  year={2010},
  publisher={IOP Publishing}
}

@article{hastie87a,
  title={Stability of ideal and resistive internal kink modes in toroidal geometry},
  author={Hastie, R J and Hender, T C and Carreras, B A and others},
  journal={The Physics of Fluids},
  volume={30},
  number={6},
  pages={1756--1766},
  year={1987},
  publisher={AIP Publishing}
}

@article{li15a,
  title={Effects of q-profiles of a weak magnetic shear on energetic ion excited q= 1 mode in tokamak plasmas},
  author={Li, Ze-Yu and Wang, Xian-Qu and Wang, Xiao-Gang},
  journal={Chinese Physics B},
  volume={25},
  number={1},
  pages={015203},
  year={2015},
  publisher={IOP Publishing}
}

@article{meng15a,
  title={Double fishbone instability excited by energetic ions with m= n= 1 in a reversed magnetic shear tokamak plasmas},
  author={Meng, Guo and Wang, Xian-Qu and Wang, Xiao Gang and others},
  journal={Physics of Plasmas},
  volume={22},
  number={9},
  year={2015},
  publisher={AIP Publishing}
}

@article{wang15a,
  title={Energetic ion beta scaling of q≳ 1 non-resonant modes in tokamak plasmas with a weak magnetic shear configuration},
  author={Wang, Xian-Qu and Wang, Xiao-Gang},
  journal={Plasma Physics and Controlled Fusion},
  volume={57},
  number={2},
  pages={025019},
  year={2015},
  publisher={IOP Publishing}
}

@article{jardin2025resistive,
	title={On resistive interchange, double tearing, and resonant and non-resonant infernal modes in spherical tokamak negative central shear discharges},
	author={Jardin, SC},
	journal={Physics of Plasmas},
	volume={32},
	number={8},
	pages = {082503},
	year={2025},
	publisher={AIP Publishing}
}

@article{zou22a,
  title={Frequency multiplication with toroidal mode number of kink/fishbone modes on a static HL-2A-like tokamak},
  author={Zou, Z H and Zhu, P and Kim, C C and others},
  journal={Plasma Science and Technology},
  volume={24},
  number={12},
  pages={124005},
  year={2022},
  publisher={IOP Publishing}
}

@article{kim08a,
  title={Impact of velocity space distribution on hybrid kinetic-magnetohydrodynamic simulation of the (1, 1) mode},
  author={Kim, Charlson C and others},
  journal={Physics of Plasmas},
  volume={15},
  number={7},
  year={2008},
  pages={072507},
  publisher={AIP Publishing}
}

@article{sovinec04a,
  title={Nonlinear magnetohydrodynamics simulation using high-order finite elements},
  author={Sovinec, Carl R and Glasser, A H and Gianakon, T A and others},
  journal={Journal of Computational Physics},
  volume={195},
  number={1},
  pages={355--386},
  year={2004},
  publisher={Elsevier}
}

@article{brennan12a,
  title={Energetic particle effects on n= 1 resistive MHD instabilities in a DIII-D hybrid discharge},
  author={Brennan, D P and Kim, C C and La Haye, R J},
  journal={Nuclear Fusion},
  volume={52},
  number={3},
  pages={033004},
  year={2012},
  publisher={IOP Publishing}
}

@book{goldston20book,
  title={Introduction to Plasma Physics},
  author={Goldston, RJ and Rutherford, PH},
  year={1995},
  publisher={CRC Press},
}

@article{Ge2023,
	title = {Multiple interactions between fishbone instabilities and internal transport barriers in EAST plasmas},
	author = {Ge, Wanling and Wang, Zheng-Xiong and Wang, Feng and Liu, Zixi and Xu, Liqing},
	journal = {Nuclear Fusion},
	volume = {63},
	number = {1},
	pages = {016007},
	year = {2022},
	publisher = {IOP Publishing},
}

@article{Wang_2014,
	year = {2014},
	month = {jul},
	publisher = {IOP Publishing},
	volume = {56},
	number = {9},
	pages = {095013},
	author = {Wang, Xian-Qu and Zhang, Rui-Bin and Qin, Liang and Wang, Xiao-Gang},
	title = {Non-resonant fishbone instabilities of qmin ≳ 1 in tokamak plasmas with weakly reversed magnetic shear},
	journal = {Plasma Physics and Controlled Fusion},
}

@article{Soltwisch_1986,
	title={Current distribution measurement in a tokamak by FIR polarimetry (invited)},
	author={Soltwisch, H.},
	journal={Review of Scientific Instruments},
	volume={57},
	number={8},
	pages={1939--1944},
	year={1986},
	publisher={AIP}
}

@article{Levinton_1993,
	title={q-profile measurements in the Tokamak Fusion Test Reactor},
	author={Levinton, F. M. and Batha, S. H. and Yamada, M. and Zarnstorff, M. C.},
	journal={Physics of Fluids B: Plasma Physics},
	volume={5},
	number={7},
	pages={2554--2561},
	year={1993},
	publisher={AIP}
}
	%Figures
	
	\clearpage

	\newpage
\begin{figure}[h]
		\centering
		\begin{minipage}{0.66\linewidth} 
		\raggedleft
		\subfigure{
			\includegraphics[width=1\linewidth]{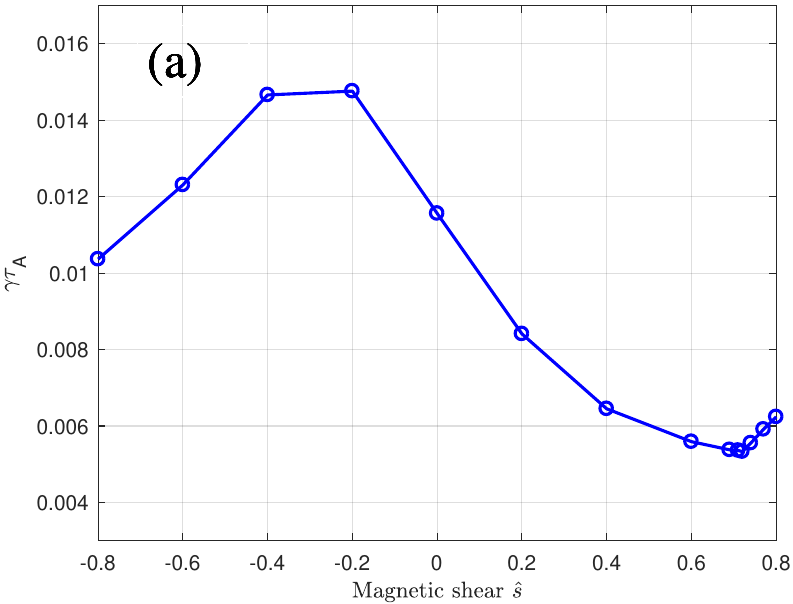}
			\label{fig:fig2.1}
		}\\
		\subfigure{
			\includegraphics[width=1\linewidth]{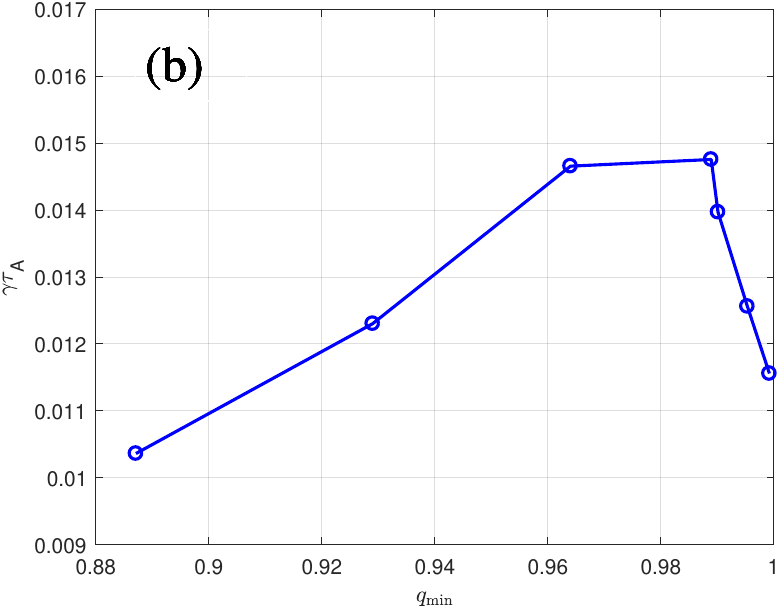}
			\label{fig:fig2.2}
		}

	\end{minipage}
	\caption{Normalized growth rates of the $(m,n)$=$(1,1)$ mode as functions (a) the magnetic shear $\hat{s}$ at $q=1$ surface and (b) the minimum of $q$ profile $q_{min}$ in absence of EPs.}
		\label{fig:fig2}
	
\end{figure}
	\clearpage
	\newpage
	
	\begin{figure}[h]
		\centering
		\subfigure{
			\includegraphics[width=0.45\linewidth]{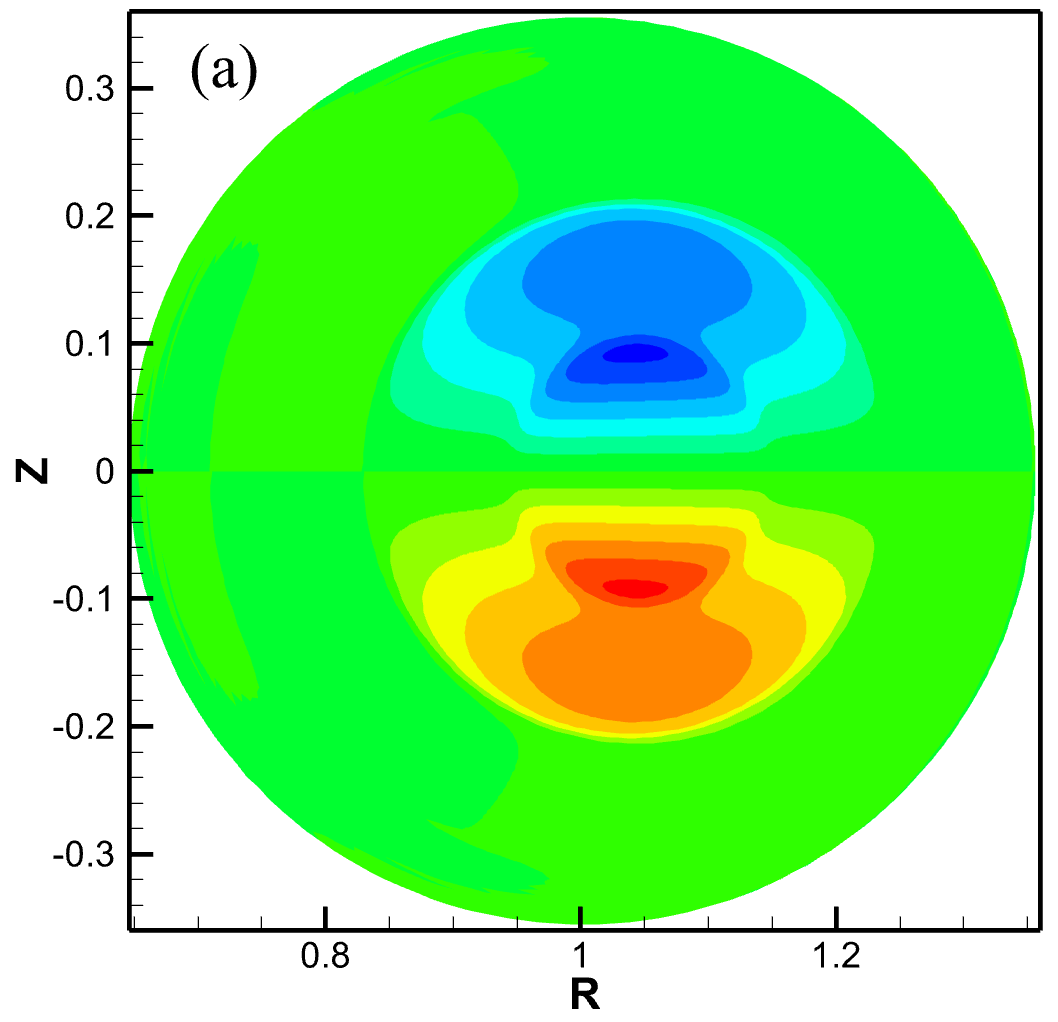}
			\label{fig:fig3.1}
		}
		\subfigure{
			\includegraphics[width=0.45\linewidth]{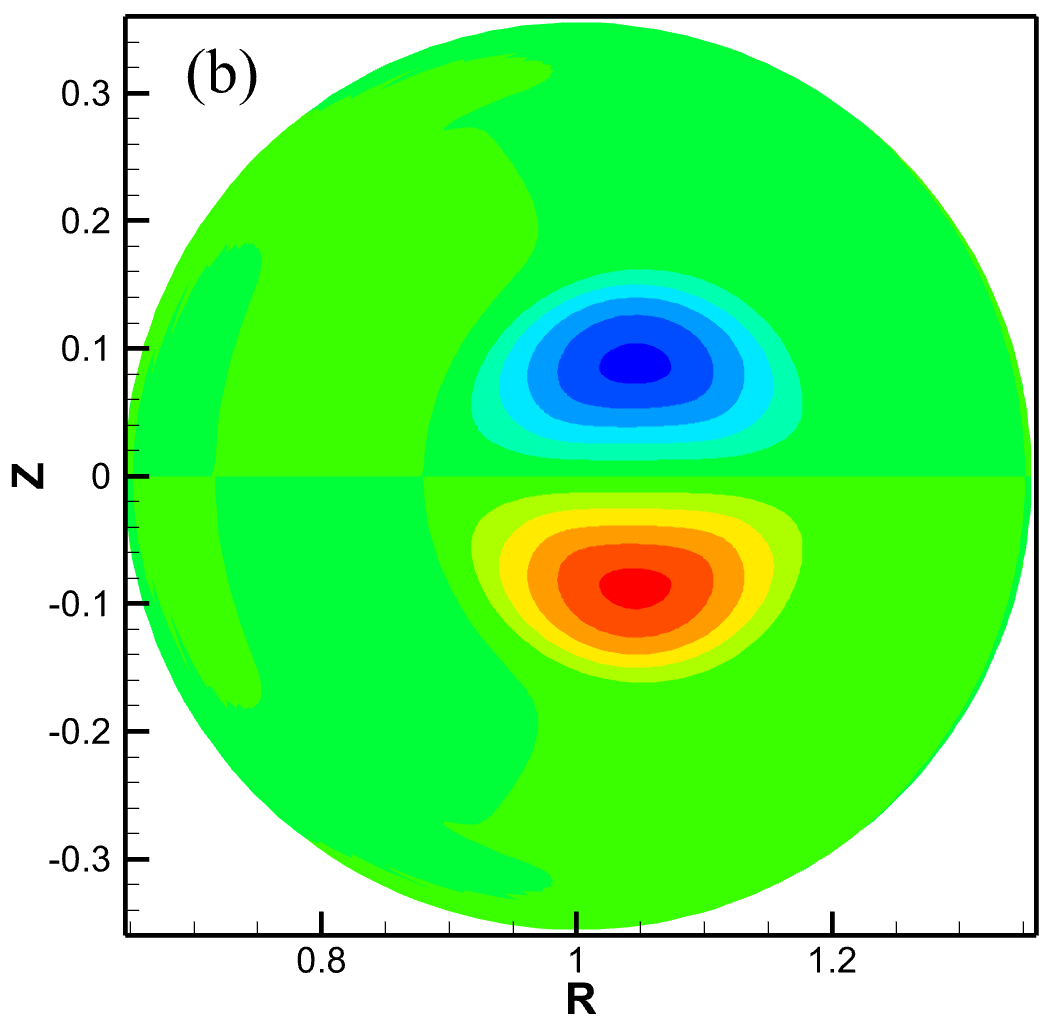}
			\label{fig:fig3.2}
		}
		\subfigure{
			\includegraphics[width=0.45\linewidth]{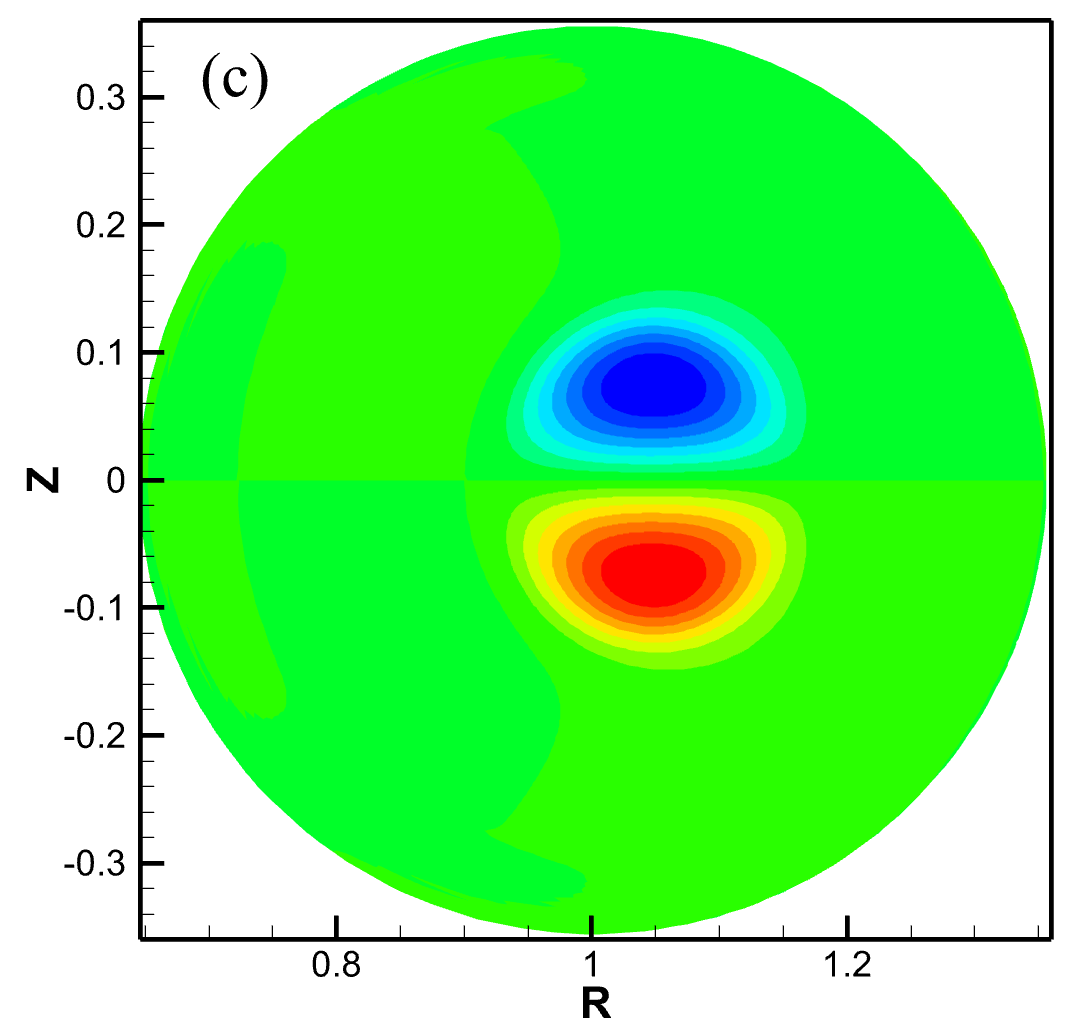}
			\label{fig:fig3.3}
		}
		\subfigure{
			\includegraphics[width=0.45\linewidth]{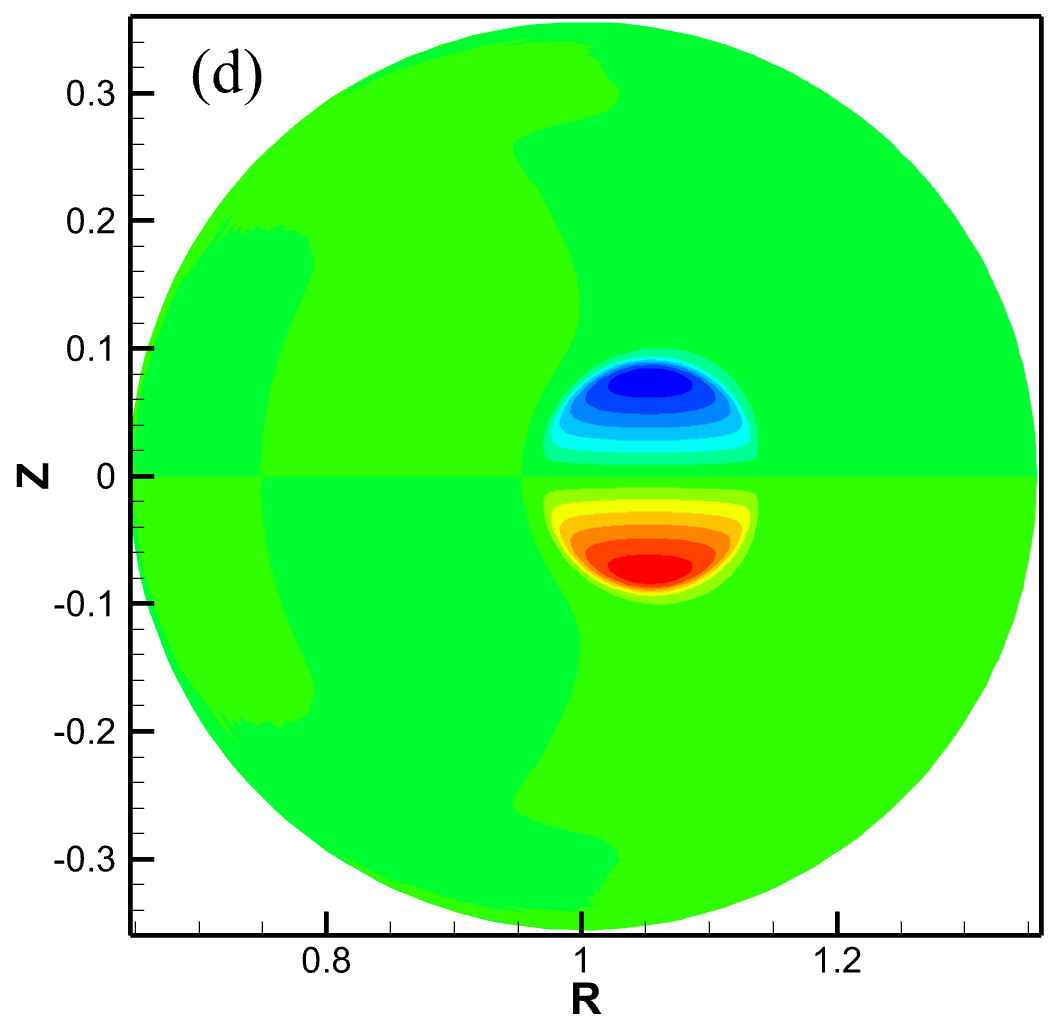}
			\label{fig:fig3.4}
		}
		\subfigure{
			\includegraphics[width=0.45\linewidth]{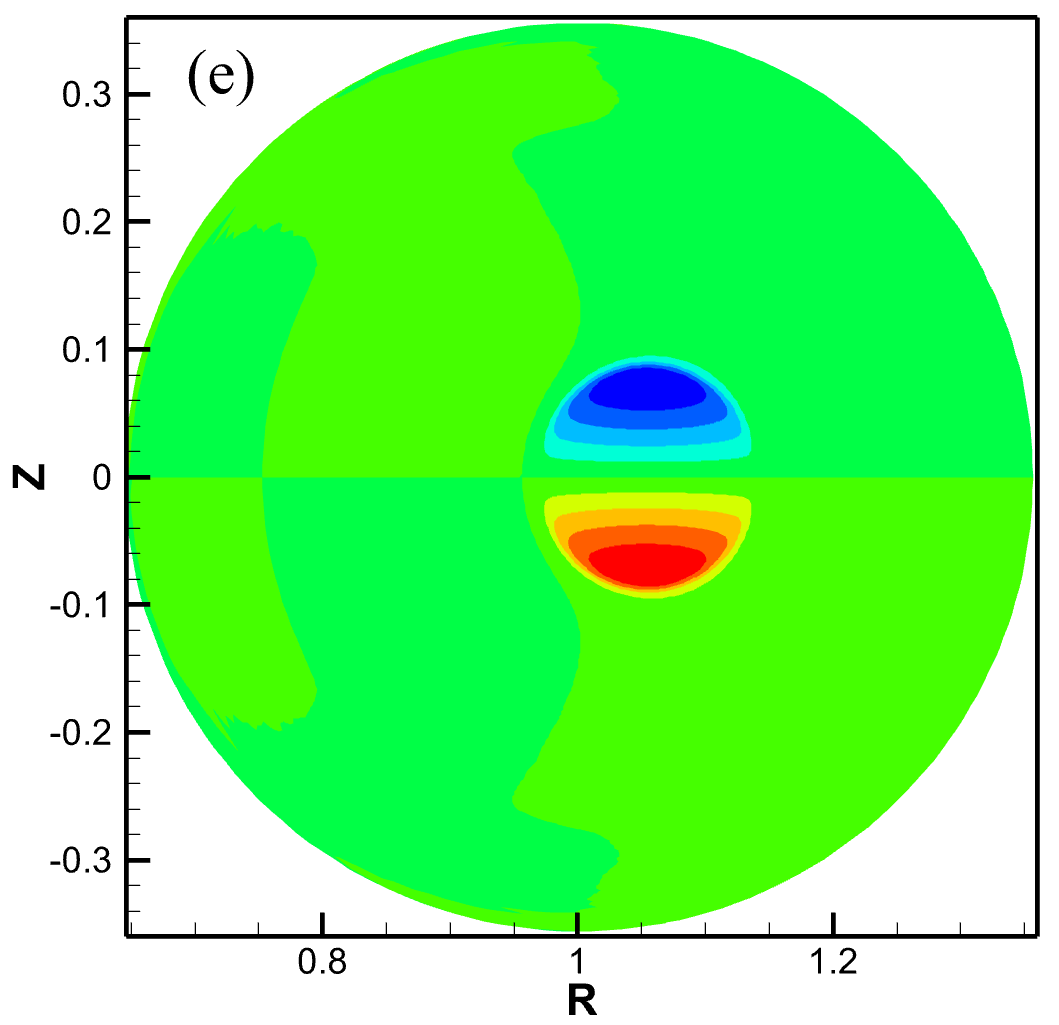}
			\label{fig:fig3.5}
		}
		\caption{Pressure perturbation contours of the $1/1$ kink mode in absence of EPs for the magnetic shear (a) $\hat s =-0.8$, (b) $\hat s = -0.2$, (c) $\hat s = 0$, (d) $\hat s = 0.7$, (e) $\hat s = 0.8$ at $q=1$ surface. }
		\label{fig:fig3}
	\end{figure}
	\clearpage

	\newpage
	
		\begin{figure}[h]
			\centering
		\begin{minipage}{0.66\linewidth} 
				\raggedleft
	
		\subfigure{
			\includegraphics[width=1\linewidth]{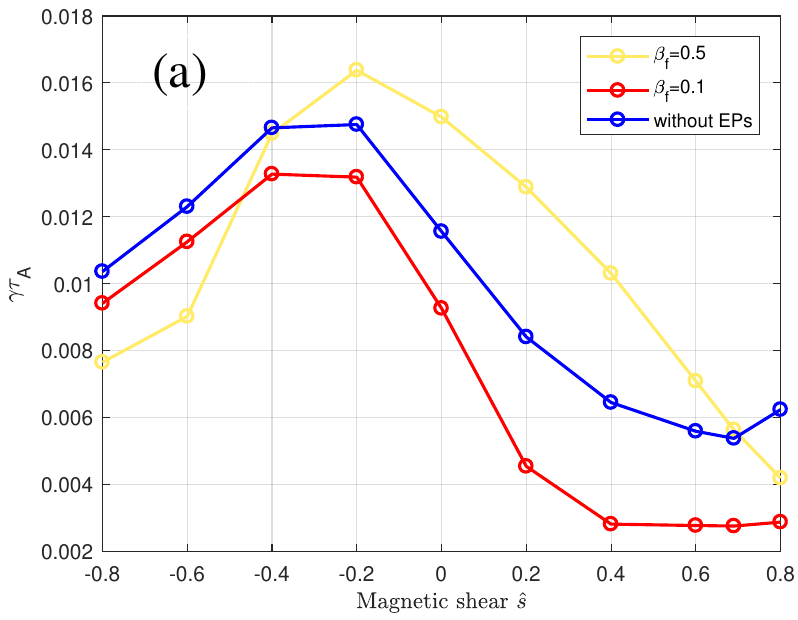}
			\label{fig:fig4.1}
		}\\
		\subfigure{
			\includegraphics[width=1\linewidth]{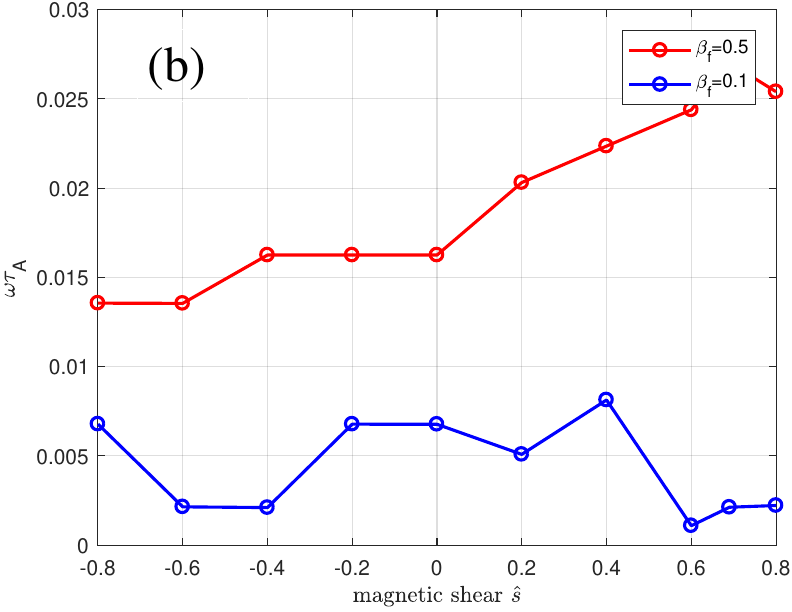}
			\label{fig:fig4.2}
		}
		\end{minipage}
		\caption{(a) Normalized growth rates and (b) real frequencies as functions of the magnetic shear $\hat{s}$ at $q=1$ surface for $\beta\rm_f=0.1, 0.5$.}
		\label{fig:fig4}
	\end{figure}
	\clearpage

	\newpage
	
	\begin{figure}[h]
		\centering
		\includegraphics[height=0.8\textheight]{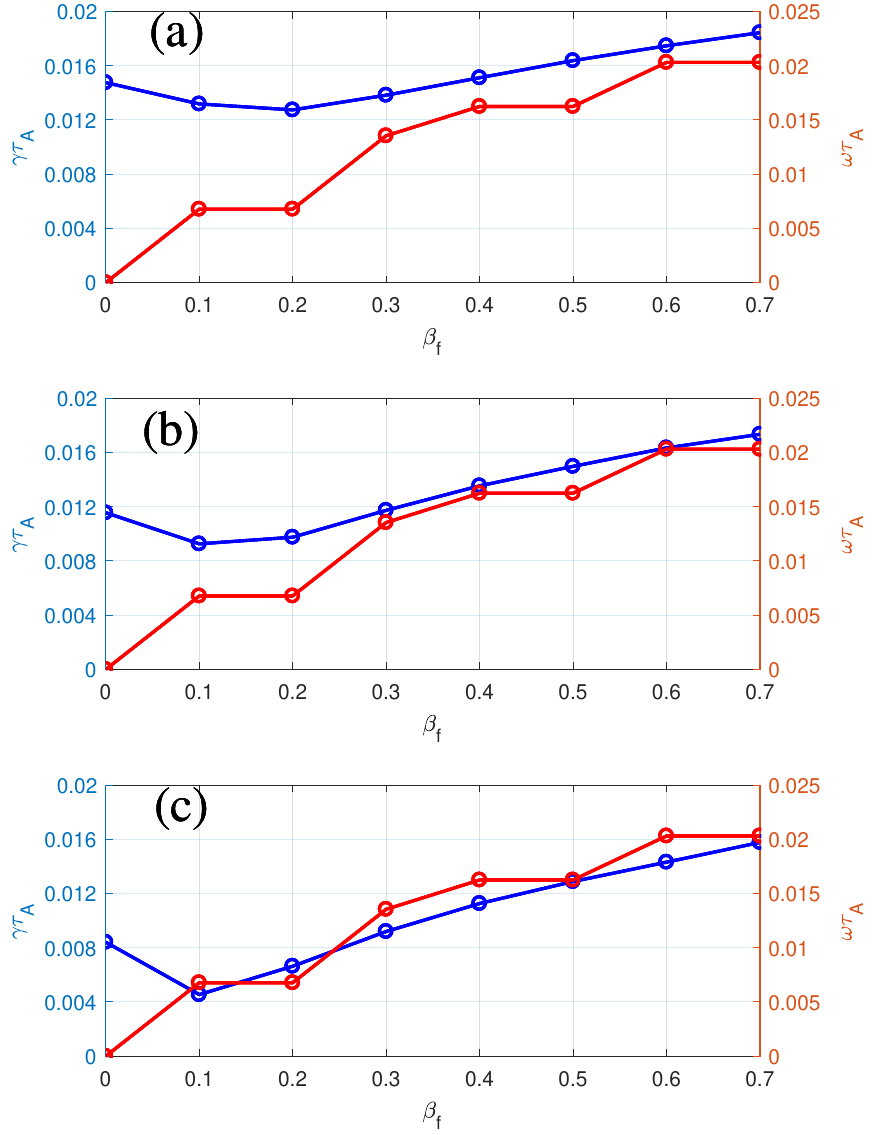}
		\caption{Normalized growth rates and real frequencies as functions of the EP beta fraction $\beta\rm_f$ for (a) $\hat{s}=-0.2$, (b) $\hat{s}=0$, (c) $\hat{s}=0.2$ at $q=1$ surface.}
		\label{fig:fig5}
	\end{figure}
	\clearpage
	
	\newpage
	
	\begin{figure}[h]
		\centering
		\includegraphics[height=0.55\textheight]{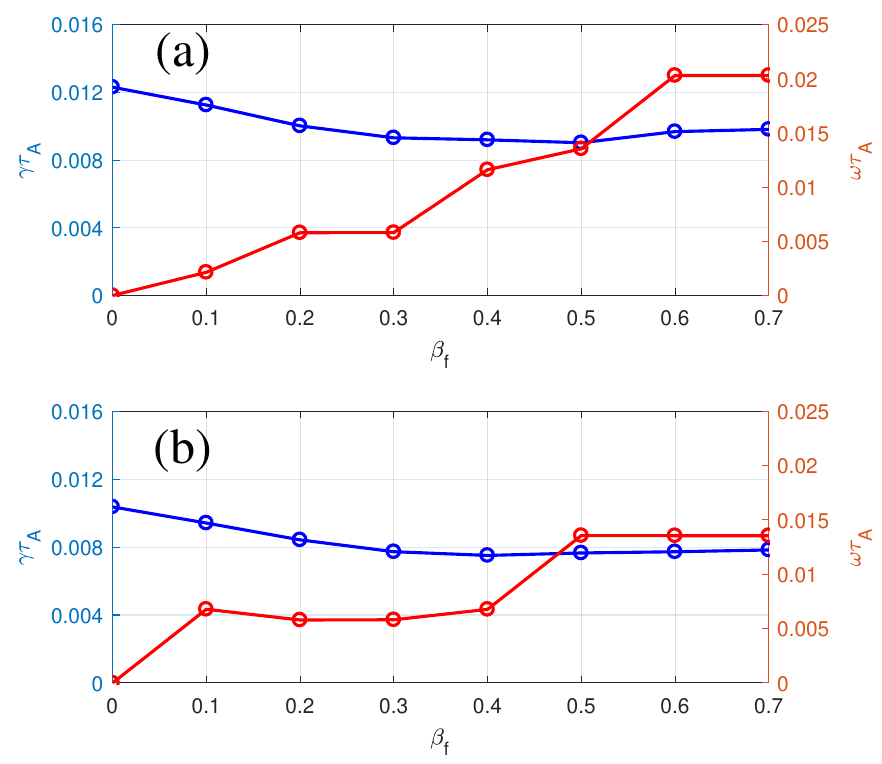}
		\caption{Normalized growth rates and real frequencies as functions of the EP beta fraction $\beta\rm_f$ for (a) $\hat{s}=-0.6$, (b) $\hat{s}=-0.8$ at $q=1$ surface.}
		\label{fig:fig6}
	\end{figure}
	
	\clearpage
	
\begin{figure}[h]
	\centering
	\subfigure{
		\includegraphics[width=0.45\linewidth]{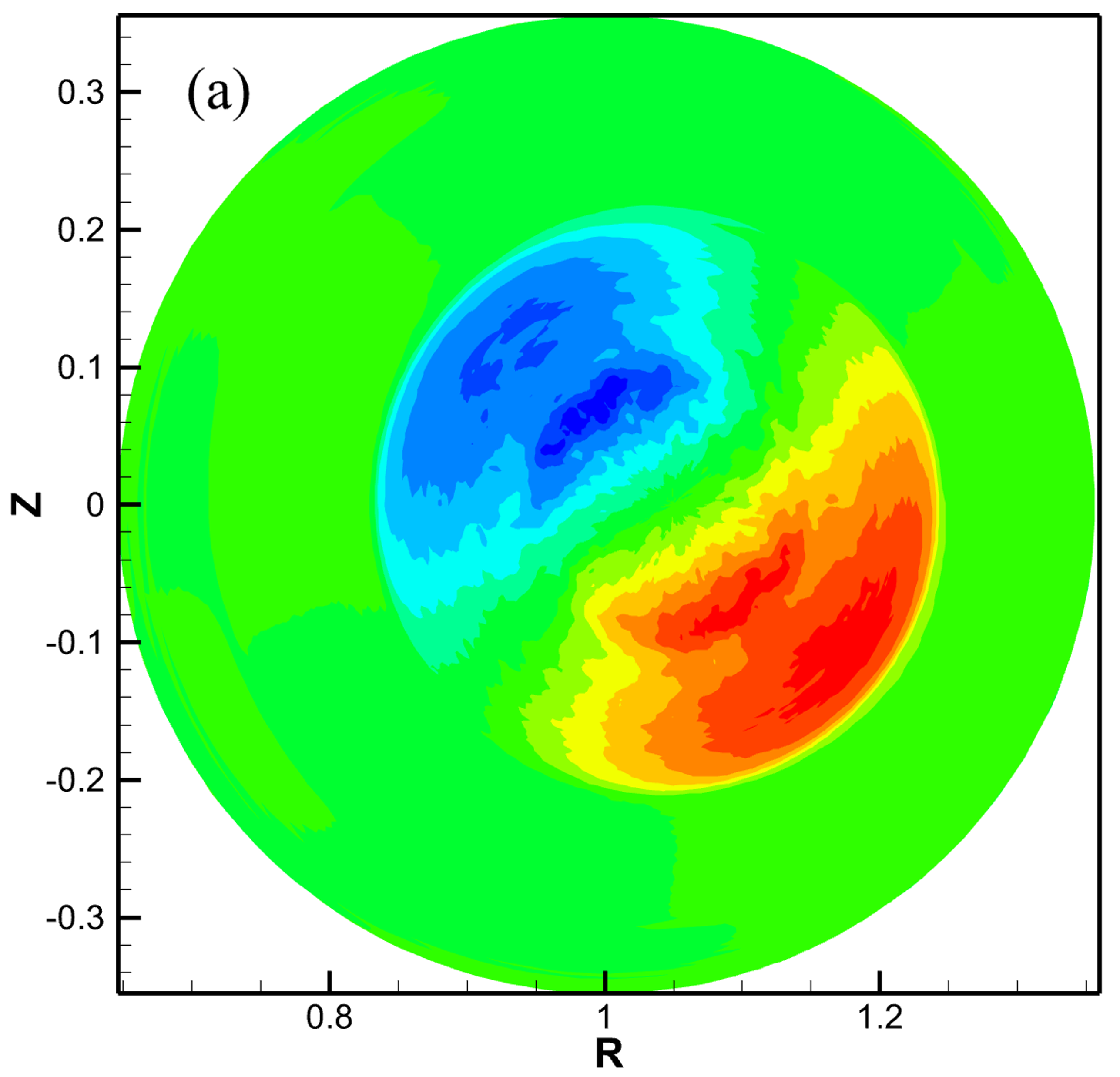}
		\label{fig:fig9.1}
	}
	\subfigure{
		\includegraphics[width=0.45\linewidth]{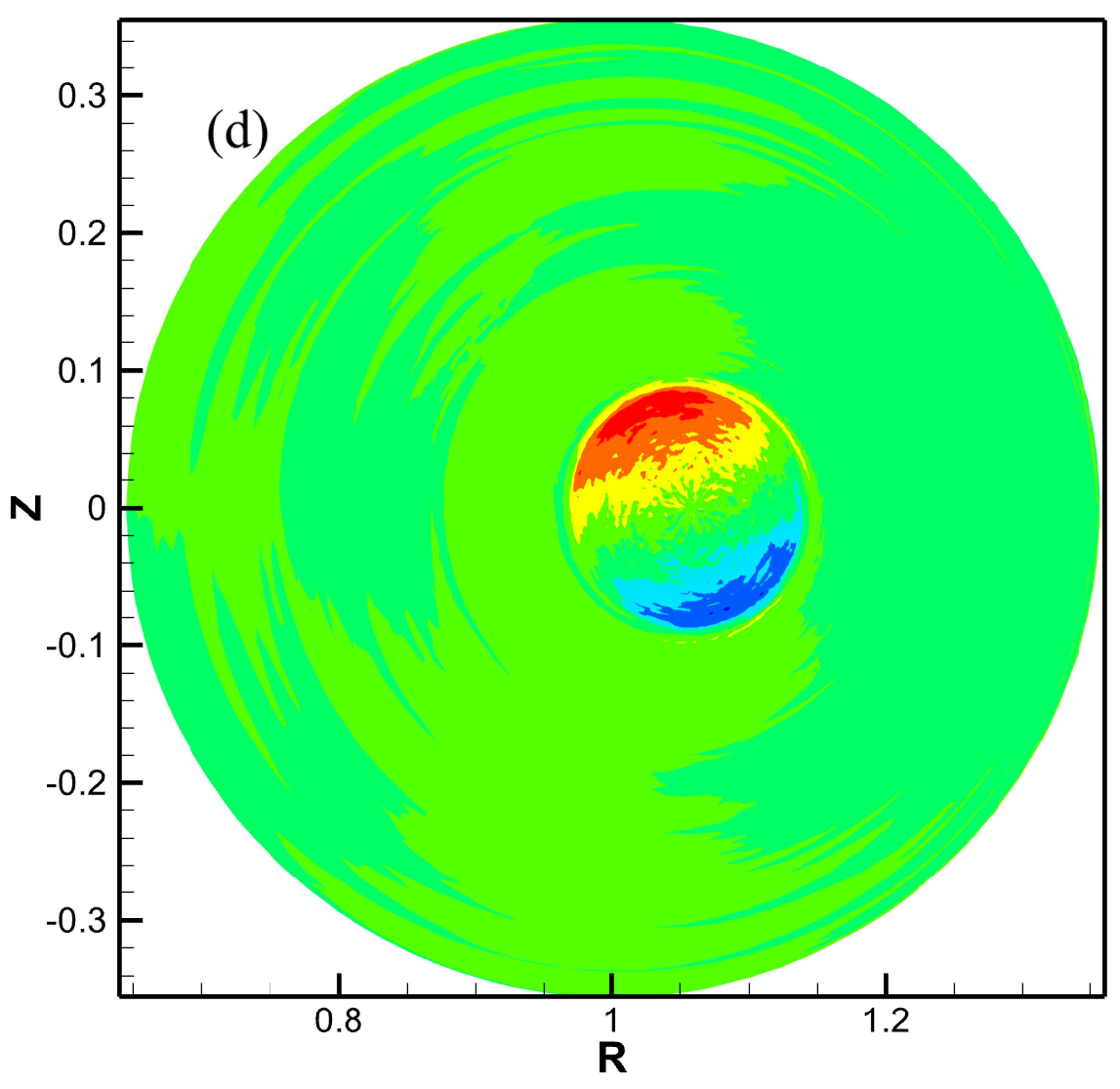}
		\label{fig:fig9.2}
	}
	\subfigure{
		\includegraphics[width=0.45\linewidth]{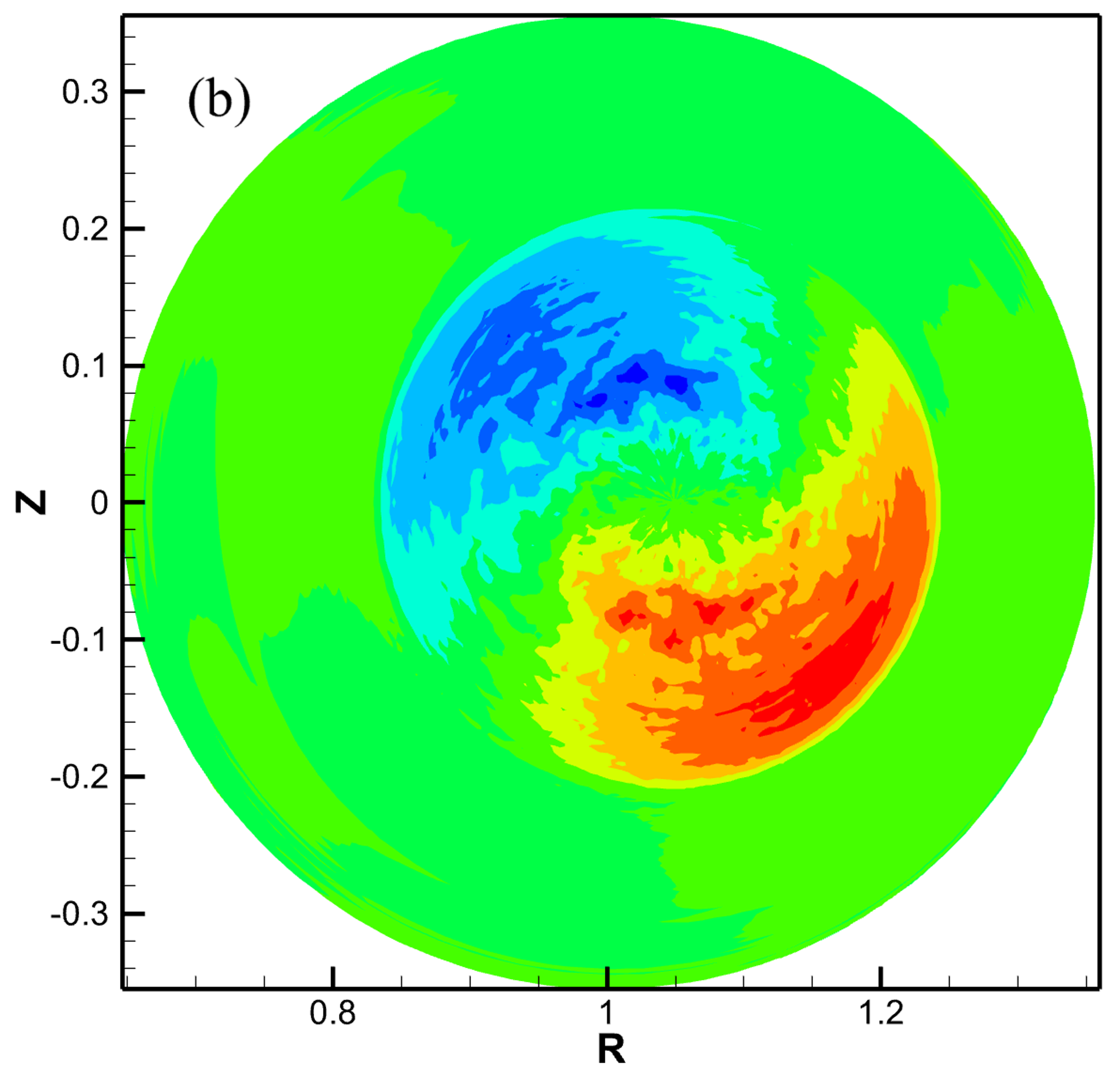}
		\label{fig:fig9.3}
	}
	\subfigure{
		\includegraphics[width=0.45\linewidth]{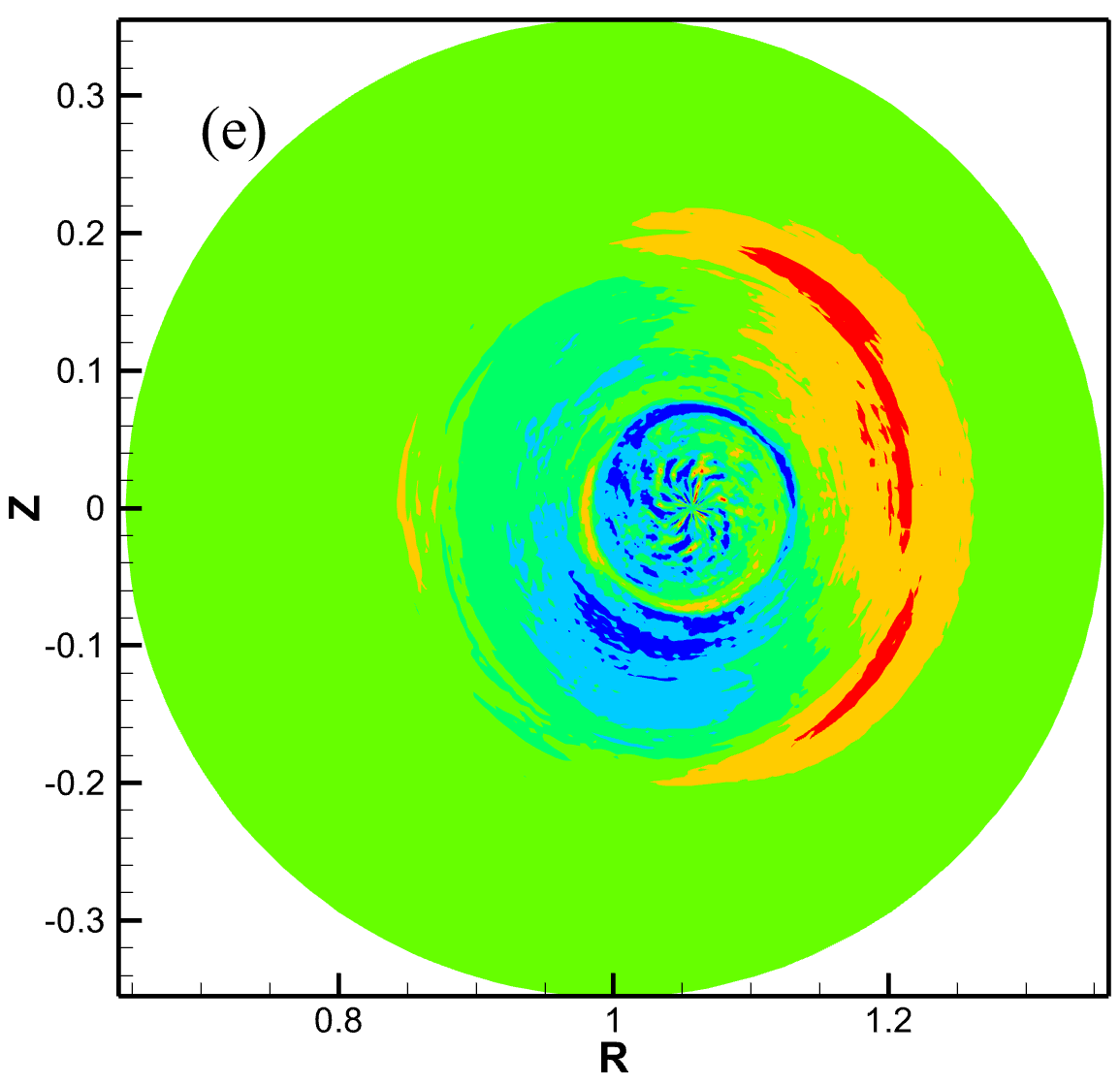}
		\label{fig:fig9.4}
	}
	\subfigure{
		\includegraphics[width=0.45\linewidth]{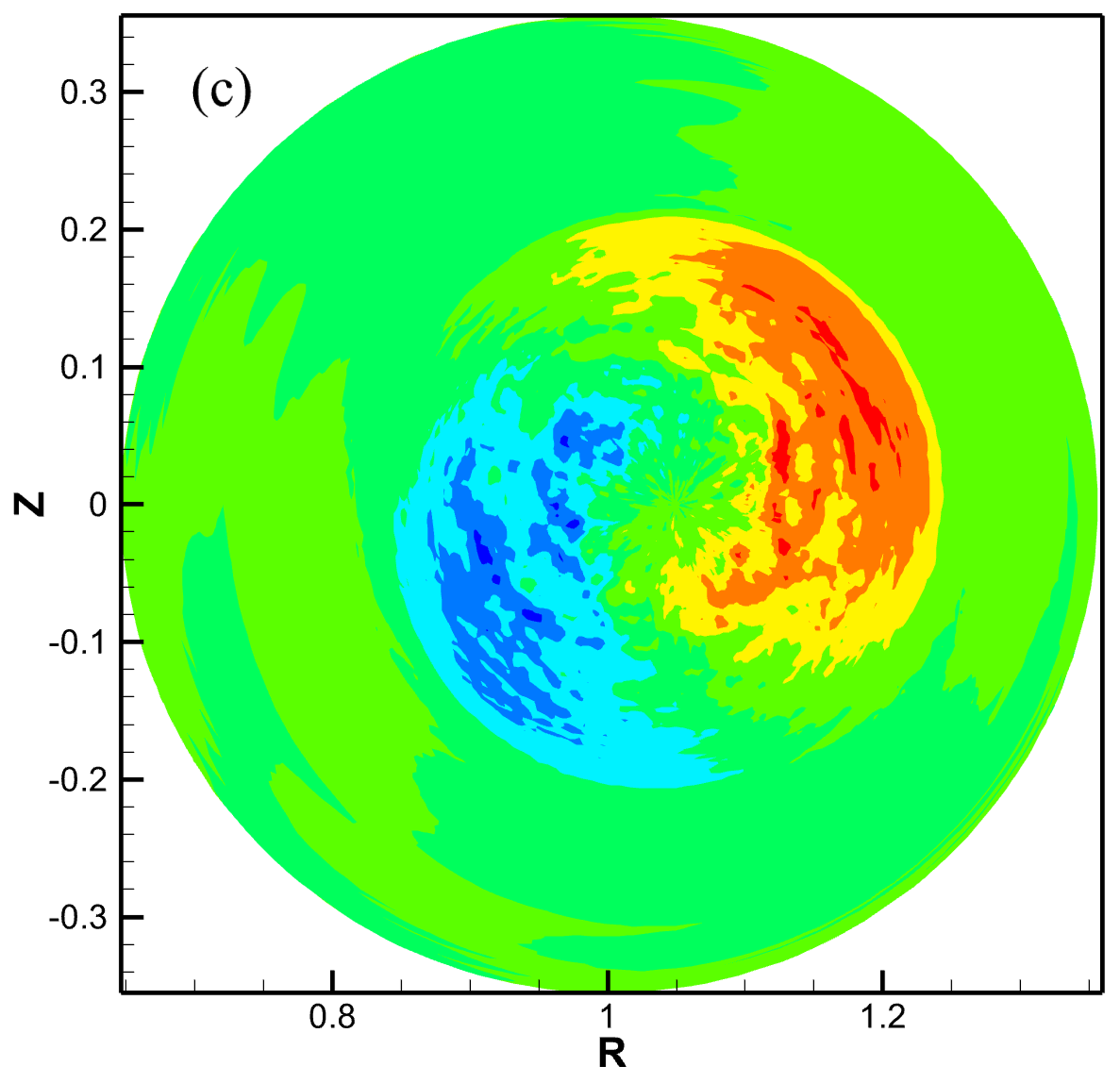}
		\label{fig:fig9.5}
	}
	\subfigure{
		\includegraphics[width=0.45\linewidth]{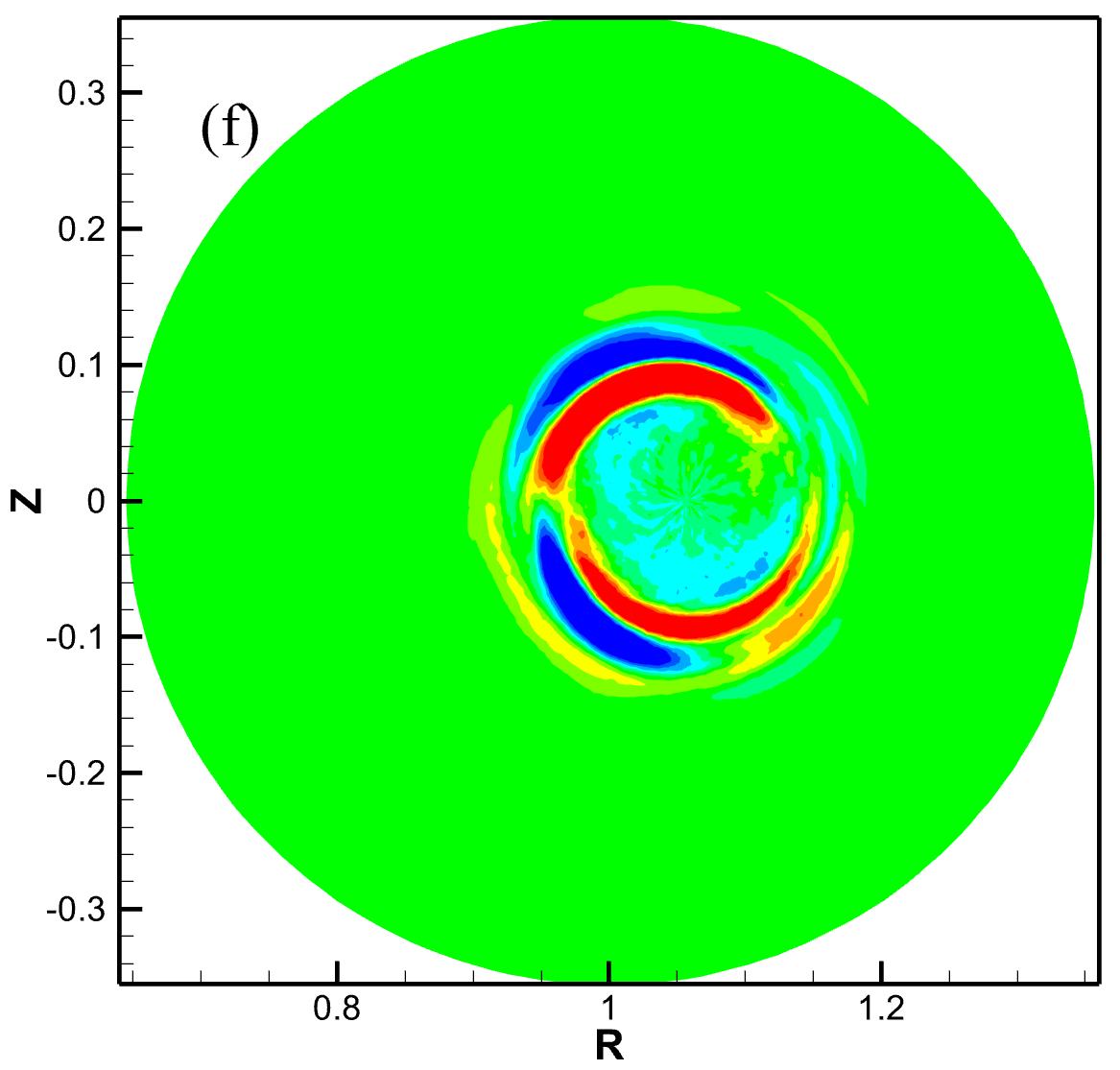}
		\label{fig:fig9.6}
	}
	\caption{Pressure perturbation contours of the $1/1$ kink mode for (a-c) $\hat{s}=-0.8$ and (d-f) $\hat{s}=0.8$ at $q=1$ surface with various EP-$\beta$ fractions: [(a) and (d)] $\beta_{\rm f}=0.2$, [(b) and (e)] $\beta_{\rm f}=0.4$, and [(c) and (f)] $\beta_{\rm f}=0.6$.}
	\label{fig:fig9}
\end{figure}
	
	\newpage
	
	\begin{figure}[h]
		\centering
		\begin{minipage}{0.66\linewidth} 
			\raggedleft 
			\subfigure{
				\includegraphics[width=1\linewidth]{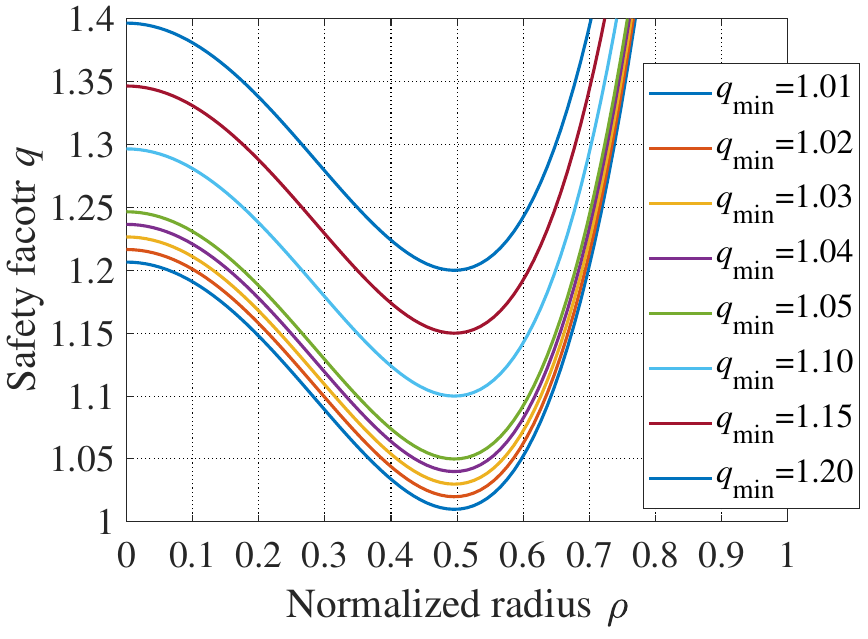}
				\label{fig:fig8.1}
			}
			\\
			
			\subfigure{
				\includegraphics[width=1\linewidth]{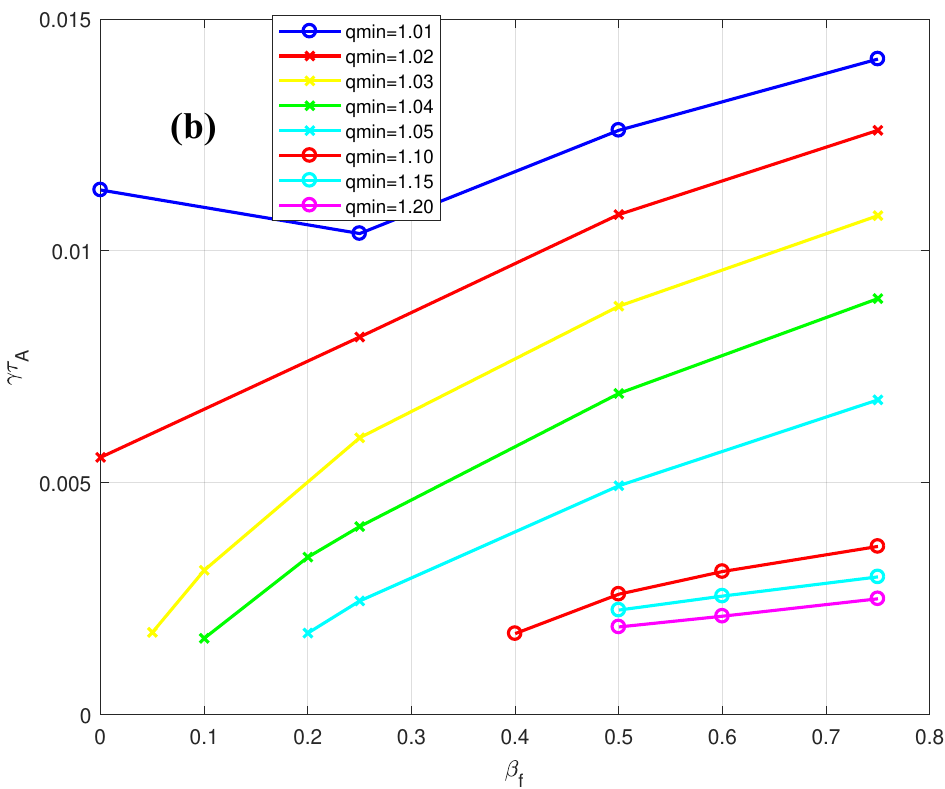}
				\label{fig:fig8.2}
			}
		\end{minipage}
		\caption{Non-resonant $q$ profiles with various $q_{\rm min}$ and (b) normalized growth rates of the non-resonant modes as functions of $\beta_{\rm f}$ for various $q_{\rm min}$.}
		\label{fig:fig8}
	\end{figure}
	
			\newpage
	
	\begin{figure}[h]
		\centering
		\begin{minipage}{0.66\linewidth} 
			\raggedleft 
		\subfigure{
			\includegraphics[width=1\linewidth]{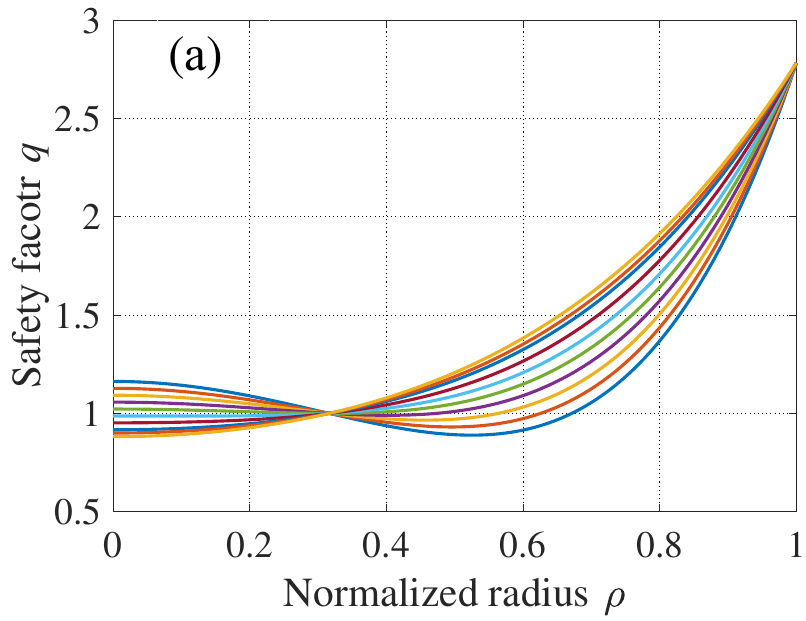}
			\label{fig:fig10.1}
		}
		\\
		
		\subfigure{
			\includegraphics[width=1\linewidth]{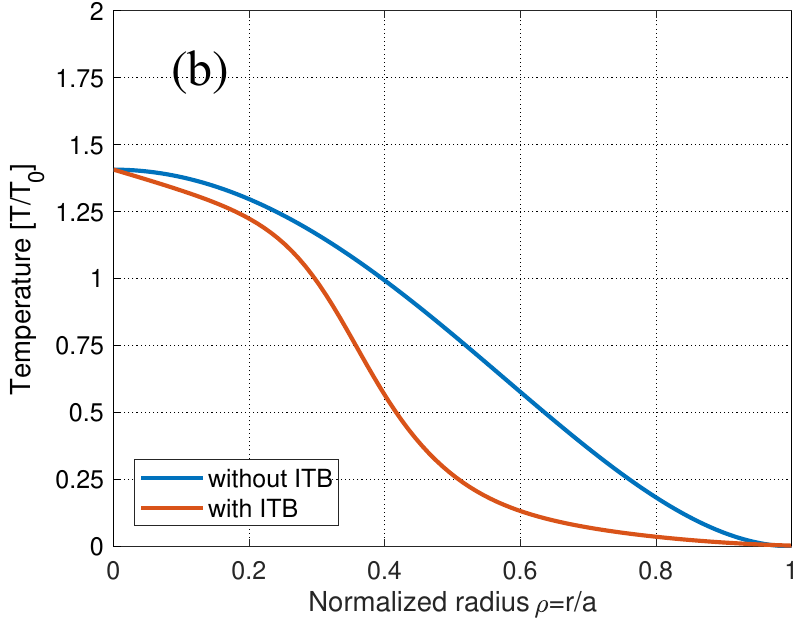}
			\label{fig:fig10.2}
		}
		\end{minipage}
		\caption{(a) $q$-profiles with various magnetic shear $\hat{s}$ at $q=1$ surface and (b) the pressure profile in absence (blue) and presence (orange) of an ITB. }
		\label{fig:fig10}
	\end{figure}
	
	\newpage
	
	\begin{figure}[h]
		\centering
		\begin{minipage}{0.66\linewidth} 
			\raggedleft 
		\subfigure{
			\includegraphics[width=1\linewidth]{image/s}
			\label{fig:fig11.1}
		}
		\subfigure{
			\includegraphics[width=1\linewidth]{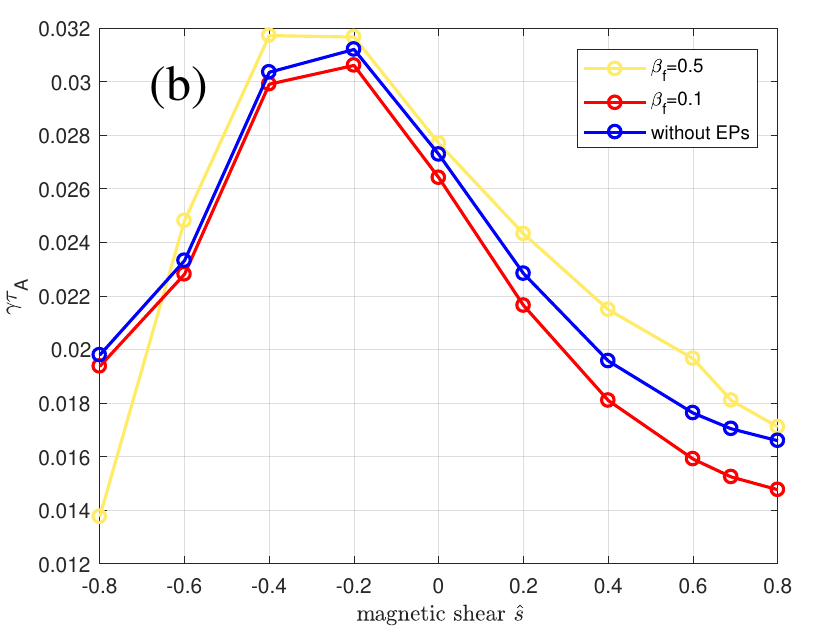}
			\label{fig:fig11.2}
		}
		\end{minipage}
		\caption{Normalized growth rates as functions of the magnetic shear $\hat{s}$ at $q=1$ surface in (a) absence and (b) presence of an ITB for various EP-$\beta$ fractions $\beta_f$.}
		\label{fig:fig11}
	\end{figure}
	
	\iffalse
	\newpage
	
	\begin{figure}[h]
		\centering
		\subfigure{
			\includegraphics[width=0.45\linewidth]{image/ITB_s-0.8_beta0}
			\label{fig:fig12.1}
		}
		\subfigure{
			\includegraphics[width=0.45\linewidth]{image/ITB_s-0.8_beta0.1}
			\label{fig:fig12.2}
		}
		\subfigure{
			\includegraphics[width=0.45\linewidth]{image/ITB_s-0.8_beta0.5}
			\label{fig:fig12.3}
		}
		\subfigure{
			\includegraphics[width=0.45\linewidth]{image/without_ITB_s-0.8_beta0}
			\label{fig:fig12.4}
		}
		\subfigure{
			\includegraphics[width=0.45\linewidth]{image/without_ITB_s-0.8_beta0.1}
			\label{fig:fig12.5}
		}
		\subfigure{
			\includegraphics[width=0.45\linewidth]{image/without_ITB_s-0.8_beta0.5}
			\label{fig:fig12.6}
		}
		\caption{The magnetic shear $\hat s=-0.8$, the contour plots of the mode pressure perturbation are shown in [(a), (b), (c)] for the case non-ITB, and in [(d), (e) , (f)] for the case with ITB. [(a) , (d)] $\beta\rm_f=0$, [(b) , (e)] $\beta\rm_f=0.1$, [(c) , (f)] $\beta\rm_f=0.5$.}
		\label{fig:fig12}
	\end{figure}
	\fi
	\newpage
	
	\begin{figure}[h]
		\centering
		\begin{minipage}{0.66\linewidth} 
			\raggedleft 
		\subfigure{
			\includegraphics[width=1\linewidth]{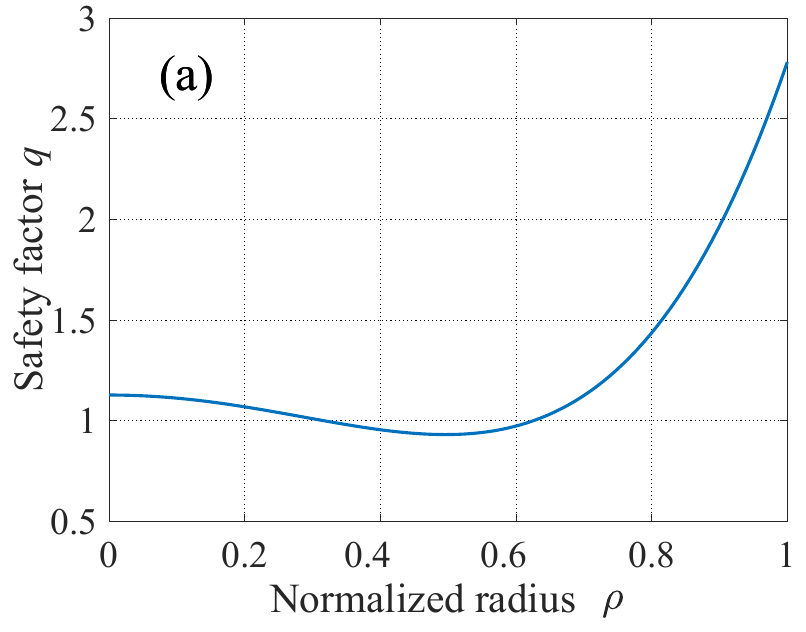}
			\label{fig:fig13.1}
		}
		\subfigure{
			\includegraphics[width=1\linewidth]{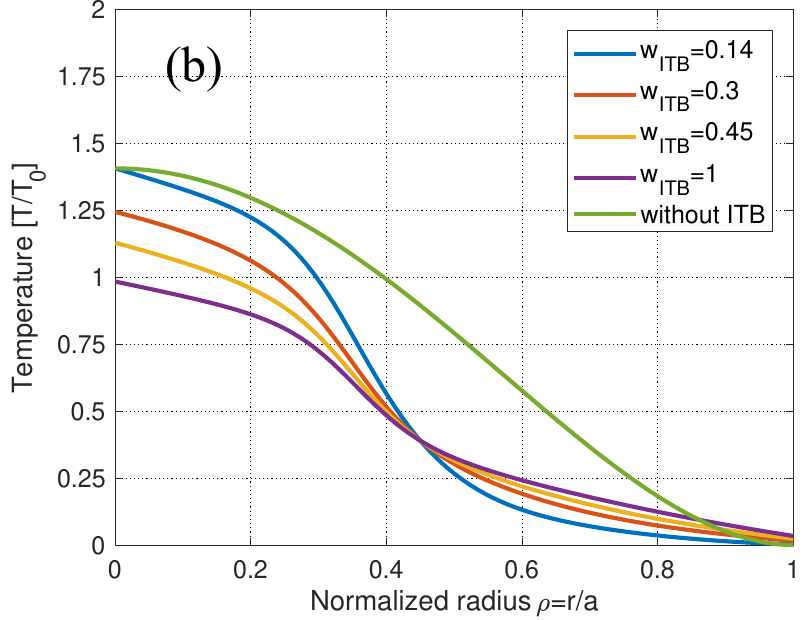}
			\label{fig:fig13.2}
		}
		\end{minipage}
		\caption{(a) $q$ profile with $\hat s=-0.6$ at $q=1$ surface, and (b) pressure profiles with various ITB widths.}
		\label{fig:fig13}
	\end{figure}
	
	\newpage
	
	\begin{figure}[h]
		\centering
		\includegraphics[width=0.7\linewidth]{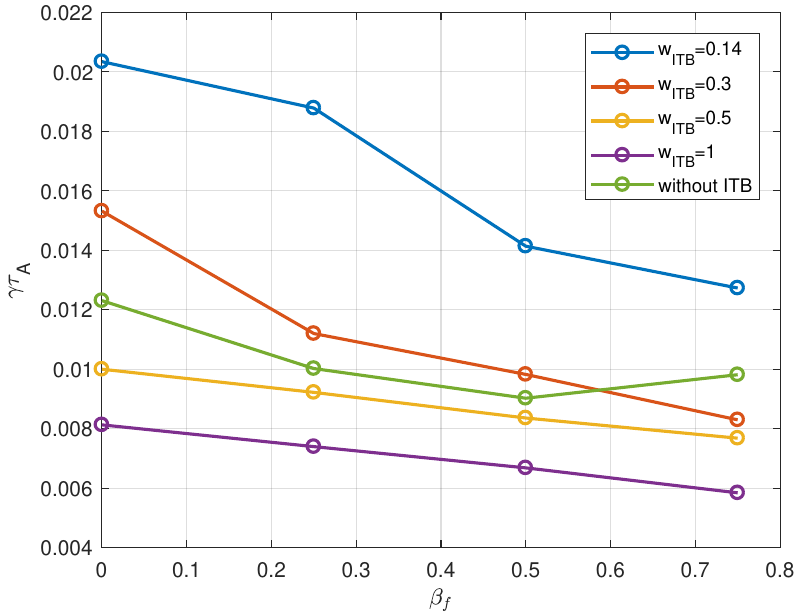}
		\caption{Normalized growth rates as functions of EP-$\beta$ fraction $\beta_{\rm f}$ for various ITB widths.}
		\label{fig:fig14}
	\end{figure}
	
		\iffalse
		
		\begin{figure}[h]
		\centering
		\subfigure{
			\includegraphics[width=0.45\linewidth]{image/HWID0.14_beta0}
			\label{fig:fig15.1}
		}
		\subfigure{
			\includegraphics[width=0.45\linewidth]{image/HWID0.3_beta0}
			\label{fig:fig15.2}
		}
		\subfigure{
			\includegraphics[width=0.45\linewidth]{image/HWID0.5_beta0}
			\label{fig:fig15.3}
		}
		\subfigure{
			\includegraphics[width=0.45\linewidth]{image/HWID0.14_beta0.5}
			\label{fig:fig15.4}
		}
		\subfigure{
			\includegraphics[width=0.45\linewidth]{image/HWID0.3_beta0.5}
			\label{fig:fig15.5}
		}
		\subfigure{
			\includegraphics[width=0.45\linewidth]{image/HWID0.5_beta0.5}
			\label{fig:fig15.6}
		}
		\caption{The magnetic shear $\hat s=-0.8$, the contour plots of the mode pressure perturbation are shown in [(a), (b), (c)] $\beta\rm_f=0$,  [(d), (e) , (f)] $\beta\rm_f=0.5$. [(a) , (d)] HWID=0.14, [(b) , (e)] HWID=0.3, [(c) , (f)] HWID=0.5.}
		\label{fig:fig15}
	\end{figure}
	
	\fi

\end{document}